\documentclass[twocolumn,aps,prl,longbibliography]{revtex4-2}

\usepackage{amsmath}
\usepackage{graphicx}
\usepackage{color}
\usepackage{ulem}
\usepackage{bm}

\usepackage[colorlinks,linkcolor=blue,citecolor=blue,urlcolor=blue]{hyperref}

\begin{document}
	
	\title{ Anomalous lepton acceleration in the radiation reaction dominated reflection regime }
		
	\author{Xiaofei Shen}\email{xfshen@pku.edu.cn; Present address: Peking University, Beijing, China}
	\affiliation{Max-Planck-Institut f\"ur Kernphysik, Saupfercheckweg 1, 69117 Heidelberg, Germany}
		\author{Yue-Yue Chen}\email{Present address: Department of Physics, Shanghai Normal University, Shanghai 200234, China}
	\author{Karen Z. Hatsagortsyan}\email{k.hatsagortsyan@mpi-hd.mpg.de}
	\affiliation{Max-Planck-Institut f\"ur Kernphysik, Saupfercheckweg 1, 69117 Heidelberg, Germany}
	\author{Christoph H. Keitel}
	\affiliation{Max-Planck-Institut f\"ur Kernphysik, Saupfercheckweg 1, 69117 Heidelberg, Germany}

	\date{\today}

\begin{abstract}

Relativistic electrons colliding with intense counterpropagating laser pulses are expected to lose energy through radiation reaction.  However, we  reveal a counterintuitive regime where reflected leptons (including incident electrons, generated electrons, and positrons) gain significant energies when a relatively loosely focused ultraintense laser interacts with counterpropagating electrons.  Because of strong radiation reaction, these particles  can be halted and reflected near the laser peak.  The subsequent asymmetric laser field then accelerates the reflected leptons to energies far exceeding their initial values.  
Using three-dimensional particle-in-cell simulations, we demonstrate the generation and acceleration of quasimonoenergetic positrons to multi-GeV energies with a high number conversion efficiency employing 10~PW-class lasers. These findings not only provide  a single-stage solution for positron creation and acceleration, but also offer a  promising alternative explanation for the origin of ultrahigh-energy cosmic rays as particularly those associated with intense fast radio bursts.

\end{abstract}

\maketitle

Laser-driven particle accelerators have attracted significant attention  because of their ability to sustain  orders of magnitude larger acceleration gradient compared to conventional radiofrequency-based systems. In particular, laser wakefield acceleration (LWFA) has achieved remarkable  success in the past decades~\cite{Esarey_2009,pukhov2002LWFA,gordienko2005LWFA,Lu2007,Steinke_2016,Gonsalves_2019,Golovanov2023}. While LWFA is favorable for electron acceleration, with longitudinally accelerating and transversely confining electrons, it presents notable challenges for positron acceleration, as the same fields that confine electrons tend to defocus positrons~\cite{Vieira_2014,Reichwein_2022}.  Despite decades of intensive research \cite{Shen2001,sugimoto2023positron,silva2021stable,silva2023positron,xu2020new,
chen2010relativistic,zhao2023terahertz,martinez2023creation,gamiz2024improved}, a simple and efficient method for positron generation and acceleration remains an outstanding challenge.

On the other hand, it is well known that a free lepton cannot gain or lose energy by interacting with a plane electromagnetic wave because of the restriction of the energy-momentum conservation law  (Lawson-Woodward theorem), or in more intuitive terms, because of the symmetry of the electron acceleration and deceleration in adjacent laser half-cycles. However, the latter strict condition can be circumvented when the interaction in the vacuum is limited in space-time via using tightly-focused or tailored laser pulses,  introducing asymmetry between the laser phase-dependent acceleration-deceleration, and giving rise to the vacuum laser acceleration (VLA) mechanism~\cite{Esarey_1995,Salamin_2002,Marceau_2012,Powell_2024,Thevenet_2015,Singh_2022,DeAndres_2024}. 
The ponderomotive scattering~\cite{Bucksbaum_1987} can significantly support the acceleration asymmetry~\cite{Hartemann_1995,Malka_1997,Mora_1998,Quesnel_1998,Stupakov_2001,He_2003}.

The symmetry of interaction of a free electron with a laser is naturally violated during the tunneling ionization when the electron instantly appears at rest in a certain laser phase after the tunneling from an atom (ion)~\cite{Moore_1999,Hu_2002}. The ionized electron in a plane wave asymptotically moves with a transverse momentum (with respect to the laser propagation direction) determined by the vector potential of the laser field at the ionization phase $\phi_i$: $p_\bot=-A(\phi_i)$, and with the corresponding longitudinal one in the relativistic case $p_{\parallel}=p_\bot^2/(2m)$~\cite{RMP_2012}, ideally gaining an energy $\Delta\varepsilon = m a_0^2/2$, being quite significant in the relativistic laser fields~\cite{Hu_2002,RMP_2012,Dodin_2003,Maltsev_2003,Gordon_2017,Yandow_2019,Yandow_2024}, e.g. reaching 12~GeV  with  state-of-the-art laser intensities  $\sim 10^{23}$~W/cm$^2$ ($a_0 \approx 220$)~\cite{Yoon_2021}. Here, $a_0=eE_0/(m\omega_0)$ is the normalized laser vector potential, with the laser field amplitude $E_0$ and frequency $\omega_0$, and the electron mass  $m$ and charge $-e$  (relativistic units  $\hbar=c=1$ are used throughout). The high acceleration gradient and simplicity of VLA have spurred significant efforts to verify and improve its performance. Nevertheless, achieving a high-quality accelerated beam via the VLA is still quite challenging. Only recently, few experiments using delicate setups to facilitate sizable injection with a plasma mirror~\cite{Thevenet_2015},  thin solid foil~\cite{Singh_2022}, or nanotip~\cite{DeAndres_2024}, were able to generate 
high electron flux though with energies limited to tens of MeV.

As the laser intensity exceeds $10^{22}$~W/cm$^2$, radiation reaction (RR) effects~\cite{RMP_2012,Fedotov_2014,Poder_2018,Cole_2018}, as well as positron generation due to nonlinear Breit-Wheeler (BW) mechanism~\cite{Ritus_1985,Burke_1997} in the laser-electron collision start to play a role, see e.g. recent works \cite{Bu_2021,Eckey_2024}. In typical setups the regime with electron gamma-factor $\gamma_e\gg a_0$ is considered,  aimed at achieving a higher quantum nonlinearity parameter $\chi_e=E'/E_{\rm cr}$~\cite{Ritus_1985}, where $E'$ is the field in the electron rest frame, and $E_{\rm cr}=m^2/e$ the QED critical field. Then, the generated photons and positrons (with  energies determined by the photon energy) move in the electron forward direction. However, in the case of  $\gamma_e\sim a_0$, an unusual, so-called reflection regime arises, wherein  the initially counterpropagating electron moves backward after the interaction due to RR~\cite{DiPiazza_2009}. The reflection regime has been employed to generate attosecond $\gamma$-rays~\cite{Li_2015,Li_2018}, or to polarize the reflected electrons~\cite{Zhuang_2023}. In principle, strong RR can halt a counterpropagating electron (positron) in the laser field, creating  dynamics similar to tunneling ionization and potentially leading to acceleration.  Yet, this regime has not been demonstrated, even numerically. 

On another front, the universe abounds with electromagnetic radiation across the spectrum from radio waves~\cite{Zhang_2023} to $\gamma$-rays~\cite{Piran_2005}.  Near their source, the radio waves can be ultrarelativistic and potentially generate ultrahigh-energy cosmic rays (UHECRs), the origin of which remains one of the most intriguing  questions in astrophysics~\cite{Letessier_2011}. However, extreme acceleration due to reflection at electromagnetic waves is unexplored.

\begin{figure}
	\begin{center}
		\includegraphics[width=0.4\textwidth]{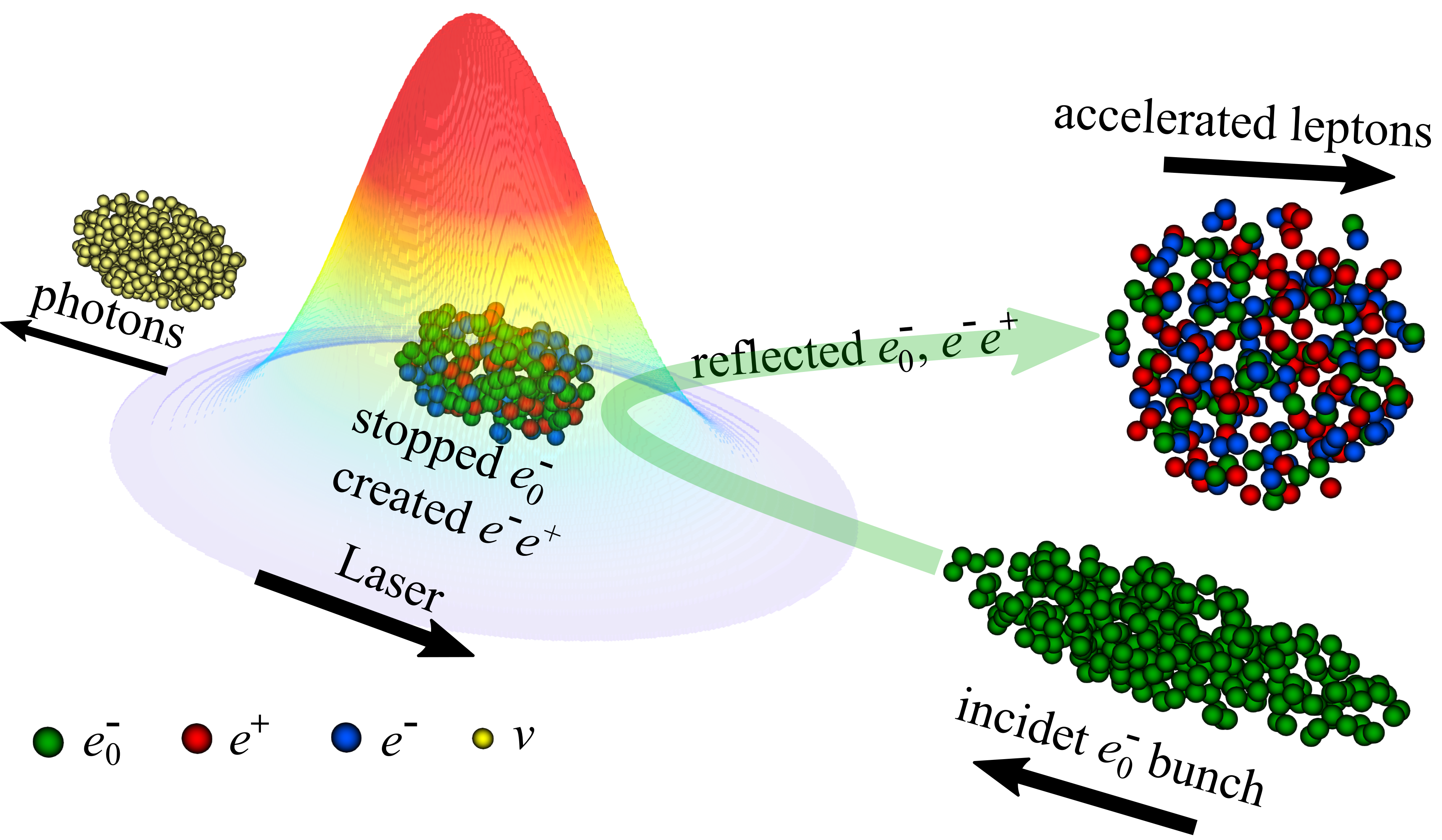}
				\caption{Schematic setup for  acceleration in the radiation reaction dominated reflection regime. A high-energy electron beam collides with an intense laser wave. Due to nonlinear Compton scattering, $\gamma$-photons  are emitted and electrons are stopped inside the wave. The emitted $\gamma$-rays decay into pairs due to nonlinear BW process. The  stopped electrons and new born pairs are further accelerated by the laser wave in the reflected direction to energies far exceeding the initial electron energy. }
						\label{fig1}
	\end{center}
\end{figure}

In this Letter, we put forward an acceleration method for leptons, employing the reflection regime in the setup of counterpropagating ultrastrong laser and relativistic electron beams in the RR dominated regime (Fig.~\ref{fig1}). Acceleration of 
the incident electrons, and created electrons and positrons via the nonlinear BW process, are discussed. Our three-dimensional (3D) particle-in-cell (PIC) simulations demonstrate that the acceleration is especially advantageous for positrons in the case of a loosely-focused laser beam. By colliding feasible LWFA electron beams with  ultraintense lasers, high-quality positron beams with energies of several GeV (about ten times higher than the initial electron energy) and high number conversion efficiency (CE) $\sim 100\%$ 
can be obtained. We explore the relevance of this scheme also for astrophysical scenarios and show the possibility of  UHECR beyond of the PeV range driven by intense fast radio bursts (FRBs).

To validate the fundamental concept of   reflection acceleration, we conducted 3D PIC simulations using the WarpX code \cite{Warp-X,Cowan_2013,Gonsalves_2019,Solver}. The laser intensity and the electron initial energy are chosen to fulfill the reflection  condition inside the laser beam $\gamma_{e0}\sim a_0$, ensure the RR dominated regime  $\alpha a_0\chi_e\gtrsim 1$~\cite{RMP_2012},  and to enable copious pair creation $ \chi_e\gtrsim 1$ ~\cite{Ritus_1985}, where $\alpha$ is the fine structure constant, and $\chi_e\sim 2\gamma_{e0}a_0\omega_0/m$. These conditions during the interaction are fulfilled when we use $a_0=550$  (the intensity of $4\times10^{23}$ W/cm$^2$), and the  initial Lorentz-factor $\gamma_{e0}=10^3$. The laser field is $x$-polarized,  and loosely focused with the  beam waist of $w_L=3.7$~$\mu$m, wavelength $\lambda_L=1~\mu$m, and pulse  duration  25 fs. The  electron beam  has a longitudinally uniform and transversely Gaussian density distribution. The beam length is $L_e=10~\mu$m, and the transverse width $\sigma= 0.7~\mu$m. The maximum electron density is $n_{e_0} = 0.016~n_c$ (the initial electron number $N_e=5\times10^8$). The initial energy spread is  5\% and angular divergence 1 mrad. These electron beam parameters are typical for LWFA \cite{Gonsalves_2019}.    The simulation box is  $60\times60\times30~\lambda_L^3$, containing $360\times360\times1496$ cells, respectively, with 64 macroparticles per cell. A moving window along the propagation direction  is employed.

\begin{figure}[b]
	\includegraphics[width=8.cm]{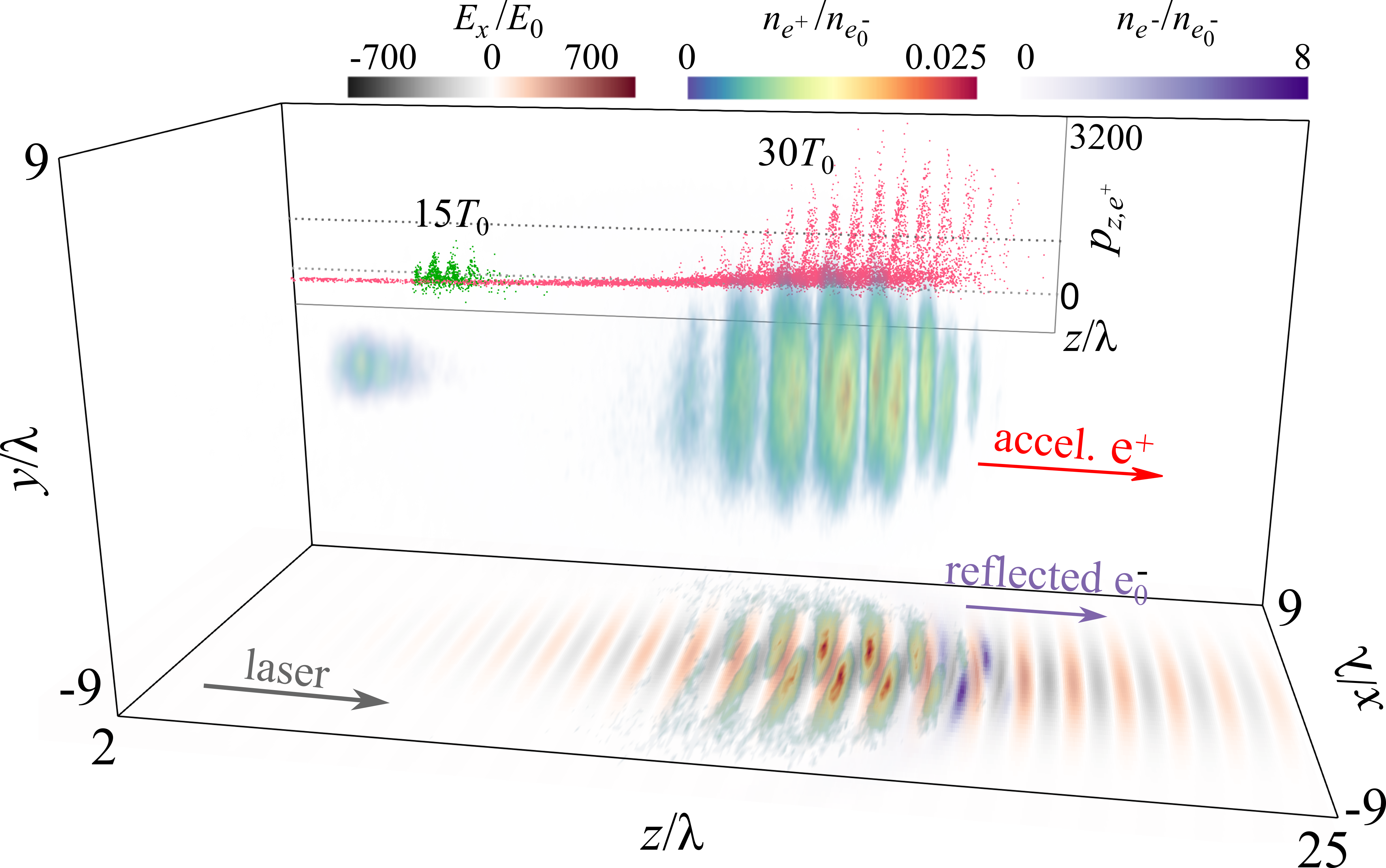}
	\caption{Creation, reflection and acceleration of positrons. Their number density distribution in space is represented by the blue-green-red colormap, while their phase spaces $(p_{z,e^+},z)$ (green dots for $t=15~T_0$ and red for $30T_0$) are displayed by the inset at the back surface (the dotted line corresponds to the electron initial momentum). The slices of the laser field and reflected electron beam density are illustrated by the gray-brown and purple colormap, respectively. 
	}\label{fig:fig2}
\end{figure}

\begin{figure}[t]
	\begin{center}
		\includegraphics[width=8.cm]{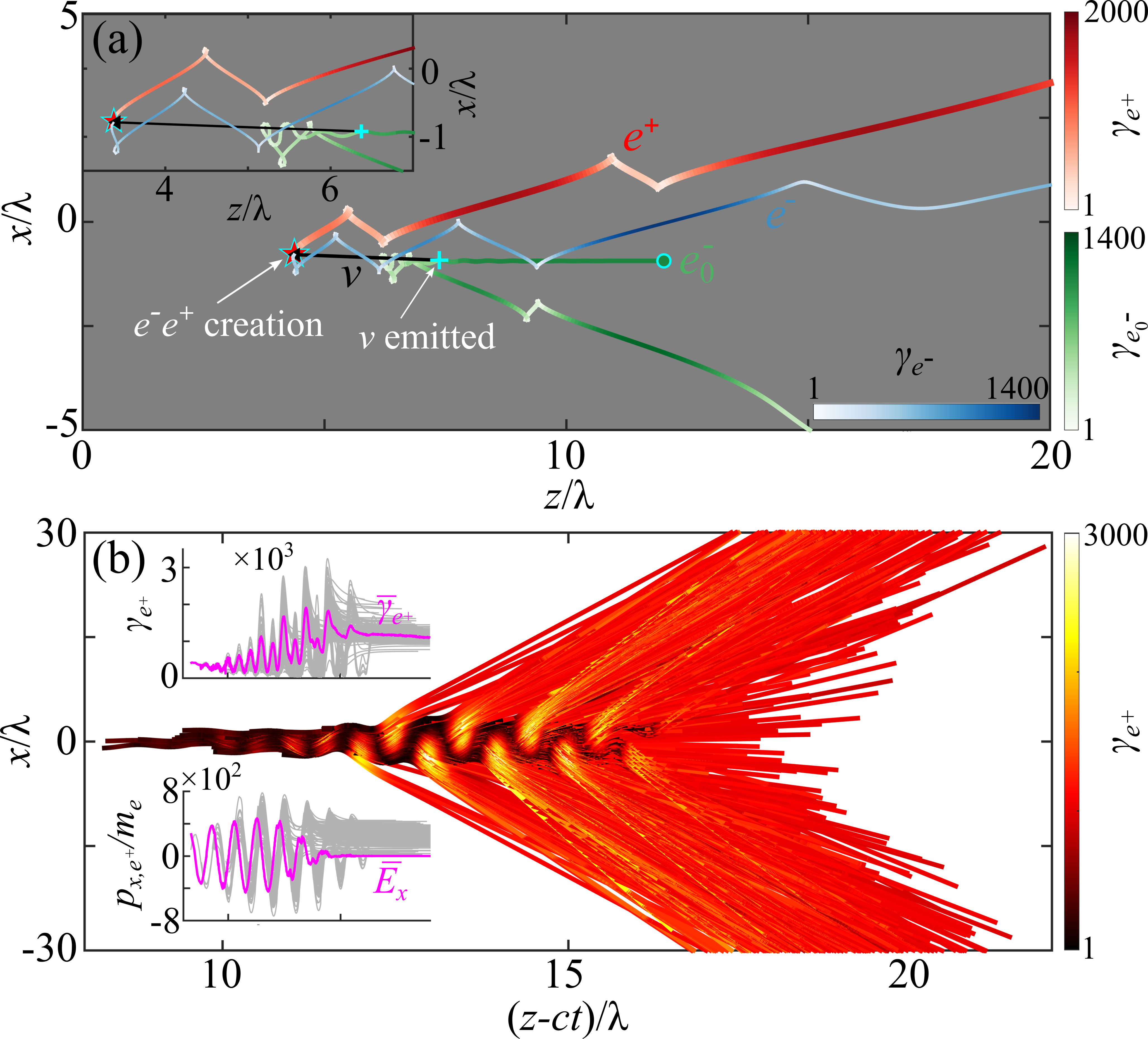}
		\caption{Particle trajectories. (a) Illustration of generation and acceleration processes of a typical pair, where the green cross and red asterisk mark the places where the photon is emitted and the pair is generated, respectively. (b) Trajectories of positrons in the comoving frame, randomly selected with $\gamma_{e^{+}}>1000$ at $t=60~T_0$. Insets show the evolution of the positrons $\gamma_{e^{+}}$ (upper) and $p_{x,e^{+}}>0$ (lower) with final $p_{x,e^{+}}>0$. The magenta line in the upper inset represents the average $\overline{\gamma}_{e^{+}}$, while  the lower shows the average $\overline{E}_x$.  
		}
		\label{fig:fig3}
	\end{center}
\end{figure}

The collision of the high-energy electrons with an intense laser beam occurs at $t=10~T_0$ ($T_0=2\pi/\omega_0$), while copious positrons begin to be created around $15~T_0$. Shortly afterward they are reflected and accelerated by the laser field. Their spatial distributions at $t=15~T_0$ and $30~T_0$ are shown in Fig.~\ref{fig:fig2}, while their phase space distributions are correspondingly presented in the inset at the back surface (green dots for $15~T_0$ and red for $30~T_0$). In Fig.~\ref{fig:fig3}(a), one representative particle is selected to illustrate the whole processes including the collision (green), photon radiation (black arrow), pair generation (red for positrons and blue for electrons) and lepton acceleration, where the emitted photon energy is 162.5~MeV with $\chi_{\nu}$ reaching 0.8. 

During the collision, a significant number of high-energy photons ( about $4\times10^7$ photons with energy above 10~MeV) is emitted via  nonlinear Compton scattering. The emitted photons move along the same direction as the initial electron beam, and further collide with the  laser pulse near the peak, leading to substantial pair production, as 
the  photon quantum strong-field parameter $\chi_\gamma\sim \chi_e\sim 1$. 
Initially, the created pairs move backwards, see the insets of Figs.~\ref{fig:fig2} and \ref{fig:fig3}(a). Then, due to the RR and Lorentz force, almost all electrons and generated pairs are reflected by the laser pulse. The positrons after the interaction mostly have nonvanishing transverse momentum $p_{x}$ correlated with the laser vector potential at the creation point, see the insets of Fig.~\ref{fig:fig3}(b). The reflected leptons are compressed in the longitudinal direction, while spreading in the transverse direction due to the unavoidable ponderomotive scattering. The peak positron density reaches about  $0.3~n_{e0}$ and then slowly decreases to be about $0.025~n_{e0}$ at $t=30~T_0$ due to the transverse expansion, as shown in Fig.~\ref{fig:fig2}.

\begin{figure}
	\includegraphics[width=0.46\textwidth]{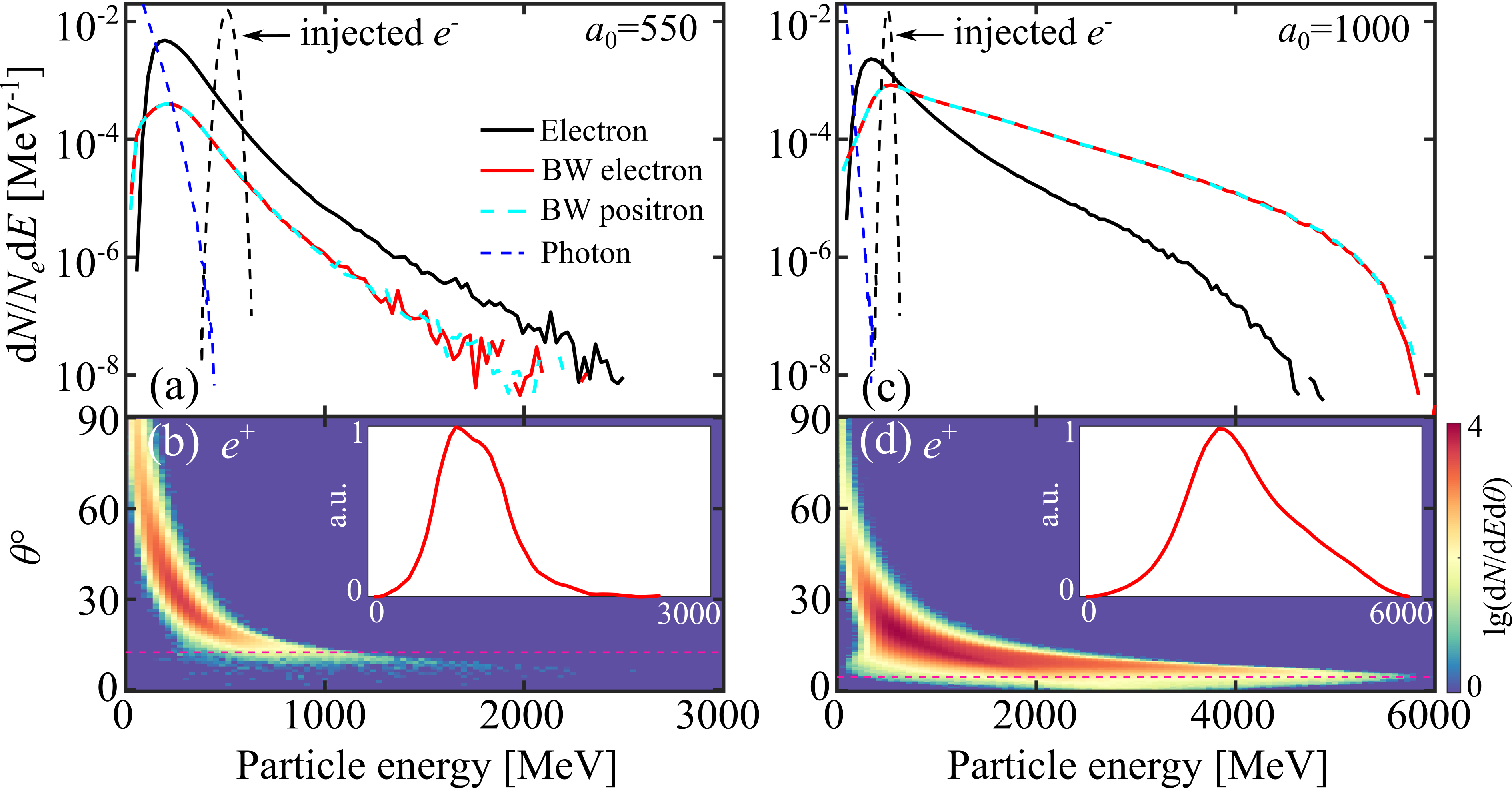}
	\caption{Energy spectra [(a), (c),  normalized to $N_e$] and angular distribution [(b), (d)] obtained with $a_0=550$ [(a), (b)] and $1000$ [(c), (d)]. Here particles are collected at simulation boundaries. Insets in (b,d) present the energy spectra of positrons selected within 12.5$^\circ$ and $4.5^\circ$, respectively. The black dashed lines in (a,c) show the energy spectrum of injected electrons. 	}
	\label{fig:fig4}
\end{figure}

While here we consider a relatively loosely focused laser pulse, the dephasing time of the accelerated leptons is much longer than the time it takes for them to exit the focal volume transversely, see End Matter (EM). As a result, the acceleration primarily ceases when the leptons leave the focal volume [Fig.~\ref{fig:fig2}(b)], and their final energies can be lower than the theoretical prediction with a plane wave. In Fig.~\ref{fig:fig4}(a), the black line represents the final electron energy distribution after leaving the laser focal spot.  The net maximum energy reaches about 2.5~GeV, about five times higher than the injected electron energy (black dashed line). With a larger  focal spot size such as $r_L=3~\mu$m, the maximum energy can be increased to about 3.55~GeV. The energy spectra of the injected electrons, created electrons and positrons are similar and their maximum energies are comparable. The beam charge of the generated positrons is about $10$~pC, leading to a number conversion efficiency (CE) from electron to positron about $12.5\%$. If we select positrons within $12.5^\circ$, a quasimonoenergetic beam with peak energy of about 0.82~GeV can be obtained, see the inset of Fig.~\ref{fig:fig4}(b).

The CE can be significantly improved by operating this regime at larger $\chi_{e}$ with an initially higher energy electron beam or laser $a_0$. For instance,  in Figs.~\ref{fig:fig4}(c,d), we present the results with $a_0=1000$, in which the positron beam charge is about 68~pC and the CE reaches about $85\%$. This means we can obtain nearly an equivalent number of pairs with the initial electrons, and it can be further enhanced with higher $a_0$ or $\gamma_{e0}$. The peak pair energy reaches 2.6~GeV with maximum energy up to 6~GeV, about 12 times higher than the initial electron energy. Interestingly, the maximum energies of the pairs are larger than those of the initial electrons. This is because most electrons are stopped before reaching the peak of the laser pulse, while the emitted photons can reach the peak and generate pairs with a larger acceleration.

With the increase of the laser intensity, the RR force becomes dominant over the ponderomotive force. Particles transverse motion is strongly suppressed and then the dephasing length becomes longer. They will have a larger chance to stay in the same half laser cycle before moving out of the laser focal spot, see EM. Therefore,  more leptons will be accelerated to higher energies, leading to a flatter energy spectrum in Fig.~\ref{fig:fig4}(c) and their divergence also becomes much smaller [Fig.~\ref{fig:fig4}(d)]. We give a simple estimation about the maximum particle energy at reasonable yield obtained in a laser wave with finite spot size $w_L$ in the RR dominated regime, see EM: 
\begin{eqnarray}
	\Delta \mathcal{E}_{\rm max} \approx \pi m a_0\frac{w_L}{\lambda_L}\approx400\sqrt{\frac{P_w}{{\rm PW}}}~{\rm MeV}.
	\label{eq:dE}
\end{eqnarray}
where $P_w$ is the laser  power in units of PW. While the energy gain is lower than for  LWFA in a PW laser, it has more favorable scaling  with the laser power ($P_w^{1/2}$ vs $P_w^{1/3}$ in LWFA~\cite{Lu2007}). We highlight that this reflection regime is suitable for generating and accelerating positrons in a single stage, while in LWFA, positrons have to be injected from another source and a special setup is required to suppress defocusing during acceleration \cite{Reichwein_2022}.

\begin{figure}
	\includegraphics[width=0.46\textwidth]{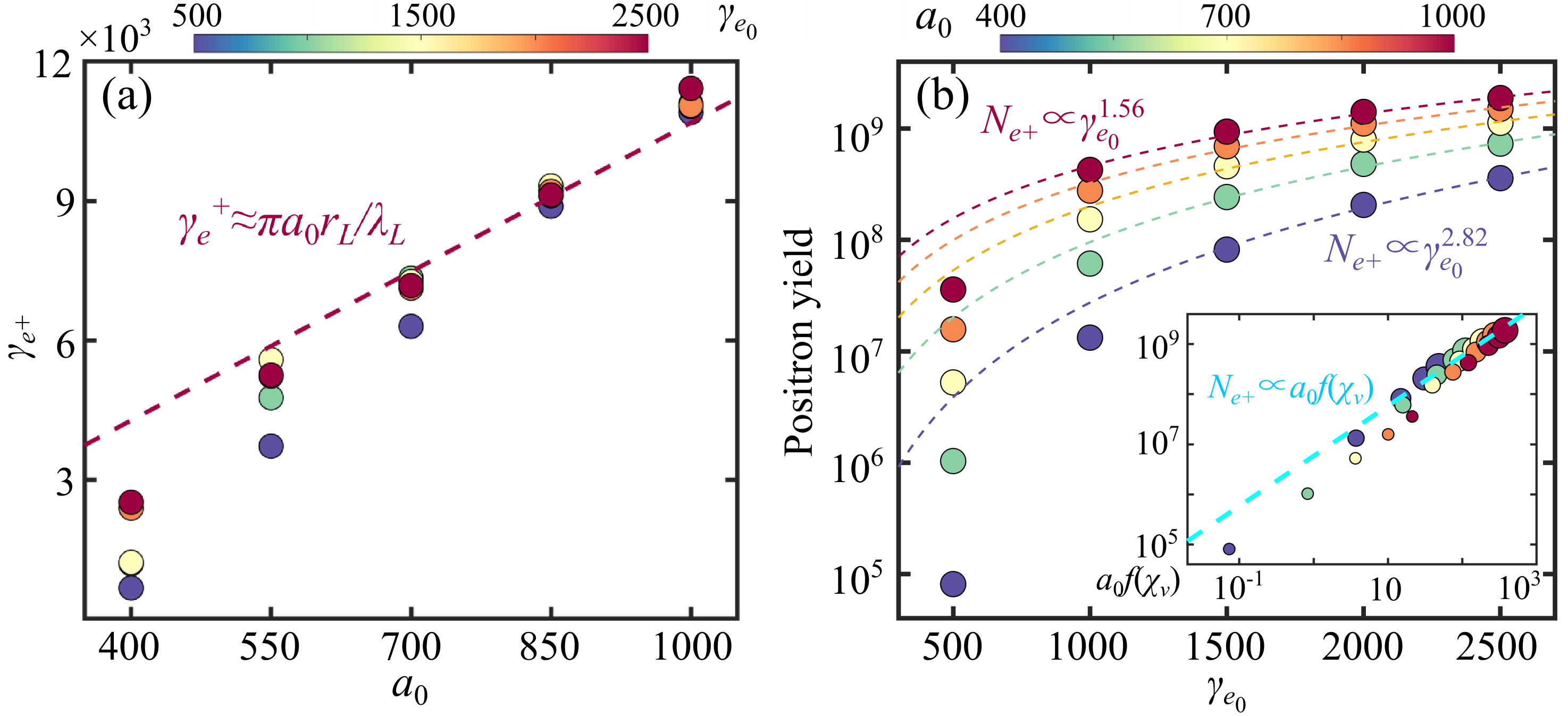}
	\caption{Scaling of the maximum positron energy $\gamma_{e^{+}}$ (a) and yield (b). In (a), the colorscale represents the initial electron $\gamma_{e0}$, while in (b) -- the parameter $a_0$. In the inset of (b), the size of the circles displays their $\gamma_{e0}$. The inset in Panel (b) demonstrates the scaling (dashed line) with respect to $a_0f(\chi_{\nu})$ given in Eq.~(\ref{eq:yield}) of End Matter. 
	}\label{fig:fig5}
\end{figure}

To validate the scaling law, we have conducted more 3D simulations by varying $a_0$ and $\gamma_{e0}$. The obtained maximum positron $\gamma_{e^{+}}$ and yield are illustrated in Figs. \ref{fig:fig5}(a,b).  One can see that the final positron energy is   proportional to $a_0$. The linear scaling in  accordance with  Eq.~(\ref{eq:dE}), takes place for $a_0\geq550$ and $\gamma_{e0}\geq1000$. In other cases,  the reflection occurs either before ($a_0>\gamma_{e0}$), or after the peak ($a_0\lesssim\gamma_{e0}$). Consequently, the fields at the moment of positron creation are smaller than $a_0$, leading to lower energies.  The positron yield [Fig.~\ref{fig:fig5}(b)] increases significantly with a higher $a_0$ and $\gamma_{e0}$. The analytical fitting formulas shown in the inset are given in EM.

Let us explore the potential of the reflection acceleration mechanism as a source of UHECR. 
In the universe, there exists strong electromagnetic radiation with frequencies spanning from radio waves to  $\gamma$-rays. For our acceleration scheme,  radio waves are preferable because they could be coherent and at the same field strength, the field parameter $a_0$  and therefore, the acceleration energy is higher than in the case of shorter wavelength. 
The brightest sources of radio waves discovered up to now are FRBs, which  recently  have attracted great attention due to their extreme luminosity varying from ${\cal L}=10^{38}~\rm erg/s$ to a few $10^{46}~\rm erg/s$ \cite{Zhang_2023,Cordes_2019,Zhang_2020}. The observed frequency is usually about $1.4~$GHz \cite{lyne2012pulsar,zhang2018FRB}, and the pulse duration spans from millisecond to tens of nanoseconds \cite{nimmo2022FRB}. The radius of FRB sources defined as $R_{s} \equiv \gamma_s\lambda_w$ is about hundreds of meters    
 \cite{kumar2017FRB}, where $\gamma_s$ is Lorentz factor of the source, and $ \lambda_w$ the wavelength of radiation. The radiation is coherent and is nearly 100$\%$ linearly polarized \cite{gajjar2018FRB_linear,lyubarsky2020FRB_reconnection}. Therefore,  FRBs can be treated as loosely focused linearly polarized electromagnetic waves \cite{popov2017GP,yang2020FRB}, which may cause reflection acceleration of particles.

Though the source of FRBs remains controversial, magnetars are at least responsible for part of them \cite{Zhang_2023}. The radius of magnetars is about 10~km, and their rotation period is about seconds, leading to a light cylinder  region of $r_{LC}\sim10^{4}$~km \cite{Lyutikov_2021}, where the last magnetic field line is closed \cite{Goldreich_1969}. While the magnetic field at the surface is extremely strong, up to $10^{13} -10^{15}$~G  \cite{katz1982magnetar,duncan1992magnetar}, it is negligible near the light cylinder, possibly about several G \cite{gralla2017quadrupole,kalapotharakos2021quadrupole,petri2015quadrupole,petri2016quadrupole}. According to the Goldreich-Juilian charge density, the local plasma density $n_{GJ}\sim 10^{12}(B/10^{12}~{\rm G})(P/0.1~{\rm s})^{-1}~{\rm cm^{-3}}\sim0.1~{\rm cm^{-3}}$ is negligible  \cite{Goldreich_1969,philippov2022pulsar_plasma}. This region is also a potential site where the FRBs are emitted \cite{zhang2022Compton,lyubarsky2020FRB_reconnection}. The wave intensity at this region propagating outwards can be roughly estimated as $I_w\sim {\cal L}/(4\pi r_{LC}^2)\lesssim 10^{20}~{\rm W/cm^2}$, corresponding to $a_0\lesssim10^7$ and $B_w\lesssim0.75~$GG.

In this region, the ultrarelativistic radio waves of FRB emerging from a magnetar \cite{kaspi2017magnetars} could encounter relativistic electrons accelerated to high energies in the magnetosphere, possibly due to  electrostatic fields \cite{lu2020FRB,wadiasingh2019FRB}, which move along the magnetic field lines towards the magnetar core. If the electron energies were $\gamma_{e0}\sim a_0$,  the reflection acceleration could work with $\chi_{e_0}\sim 1$. Based on the above parameters and our energy scaling Eq.~(\ref{eq:dE}), one can  estimate that the maximum lepton energy due to  reflection acceleration would reach the PeV range, with $w_L\sim 1$~km and $I_w\sim10^{17}~{\rm W/cm^2}$. This is still  smaller than the upper limit of the plane wave theory $40$~PeV. Thus, the reflection acceleration mechanism could be responsible, in principle, for the observed UHECR. Note that due to the synchrotron emission during the propagation  over the scale of many galaxies, leptons lose a significant amount of energy and that is why the observed lepton energy is about tens of TeV on earth~\cite{archer2018TeV_electrons,recchia2019TeV_electrons}.

While the RR dominated reflection regime does not work for ions, if one conjectures that ions are created at rest at the wave peak, they would attain energies exceeding $10^{19}- 10^{21}$~eV depending on the ions species (at the same Lorentz factor of electrons)~\cite{aab2020rays}.  This is because for $a_0\gg M/m$, with the ion mass $M$, the acceleration progress should be almost identical except the RR effects. Moreover, the acceleration distance is roughly the distance when the  intensity is decreased to half of its peak value. This is only about $10^4-10^5$~km, slightly larger than the diameter of earth, but extremely small compared to the galaxy scale. Hence, it might have a much higher probability to work compared to the Fermi acceleration which requires a Mpc (about $10^{19}$~km) acceleration distance~\cite{drury1994Rays}.

In conclusion, we put forward a  lepton acceleration mechanism which occurs when  RR effects become dominant. The counterpropagating electrons to an ultrastrong laser beam and the created secondary electrons and positrons can be stopped and reflected at the peak of the laser pulse, and then accelerated to much higher energies than the initial injection energy. While with current laser facilities, the obtained accelerated electron beam quality is difficult to compete with LWFA, the reflection acceleration provides a very simple method to create and accelerate positrons to GeV energies which remains challenging with other known schemes. The astrophysical observation data do not exclude that this acceleration mechanism could contribute to the generation of UHECR  emerging from FRBs.

\section{End Matter}

 \textit{Interaction time and dephasing.}  The electrons leave  the laser pulse either in the longitudinal (at small angles relative to the laser propagation direction) or transverse directions (at large angles). For ultrarelativistic electrons with divergence  $\theta$, the longitudinal interaction time with the focused laser beam can be estimated as $t_\parallel\sim z_R/(c\cos\theta)$, where $z_R=\pi w_L^2/\lambda_L$ is the Rayleigh length, assuming that the field is significantly damped at $z>z_R$.   The transverse interaction time can be estimated as $t_\bot\sim w_L\sqrt{1+z^2/z_R^2}/(c\sin\theta)\sim \sqrt{2}w_L/(c\sin\theta)$.  The time to slip from the laser pulse in the longitudinal direction is $t_{\parallel}^{\rm pulse}=\tau_L/[2(1-\beta_e \cos\theta)]\approx\tau_L\gamma_e^2/(1+\gamma_e^2\theta^2)$ at $\theta\ll 1$, which  is much larger than $t_\parallel$ and $t_\bot $ for our parameters for the given pulse duration $\tau_L$.

 Thus, $t_\parallel/t_\bot\sim \pi w_L\tan(\theta)/(\sqrt{2}\lambda_L)$, i.e., at small angles $\theta <\sqrt{2}\lambda_L/(\pi w_L)\ll 1$, the electron leaves the focal volume longitudinally, and vice verse. 
 
The dephasing time in a plane wave approximation is  $t_{dp}=\pi/[\omega(1-\beta_e \cos\theta)]\approx(\lambda_L/c)\gamma_e^2/(1+\gamma_e^2\theta^2)$, which in the focused laser pulse decreases to $t_{dp}\sim z_R/c$ at small angles $\theta \ll\sqrt{2}\lambda_L/(\pi w_L)=\sqrt{2}w_L/z_R$, see Ref.~\cite{Maltsev_2003}. Thus, at small angles the interaction and dephasing times coincides, which then provides the acceleration time $t_{\rm acc}\sim z_R/c$.

At large angles $\theta >\sqrt{2}\lambda_L/(\pi w_L)$, the dephasing time $t_{dp} \sim  (\lambda_L/c)\gamma_e^2/(1+\gamma_e^2\theta^2)\sim \lambda_L /(c\theta^2)$ (for $\theta\gtrsim \gamma_e^{-1}$), with $\theta\sim \sqrt{2} w_L/z_R$, and we  derive $t_{dp}\sim (\pi/2) (z_R/c)$, which also coincides with the interaction time in this case  $t_\bot\sim  \sqrt{2}w_L/(c\sin\theta)\sim t_{\rm acc}$. 

Based on the above discussion,  the acceleration time is
\begin{eqnarray}	
t_{\rm acc} \approx \left\{
	\begin{aligned}
		\frac{z_R}{c},\;\;\;\;\; \theta\ll\frac{\sqrt{2}\lambda_L}{\pi w_L}\\
		\frac{w_L}{c\theta},\;\;\;\;\;\;\; \theta>\frac{\lambda_L}{\pi w_L}\\
	\end{aligned}
	\right.
\end{eqnarray}

\textit{Acceleration scaling.} Now we can estimate the energy gain.  At small angles $\theta \ll \sqrt{2}\lambda_L/(\pi w_L)$, the energy gain is:
\begin{eqnarray}
\label{eq:dEE}
\Delta \varepsilon \sim eE_0 v_\bot t_{\parallel}\sim e E_0 z_R \theta\sim \pi m a_0 w_L/\lambda_L,
\end{eqnarray}
which is in accordance with Ref.~\cite{Maltsev_2003} and presented in Eq.~(\ref{eq:dE}). We have assumed therein an average field strength over the optical cycle of $E_0/2$, wher at large angles $\theta > \sqrt{2}\lambda_L/(\pi w_L)$, leading to $\Delta \varepsilon \sim eE_0 v_\bot t_\bot\sim eE_0 w_L$, similar to Eq.~(\ref{eq:dEE}).

\textit{Positron yield.} For an order of magnitude estimate of the positron yield, we   assume that all  electrons are reflected. According to  momentum conservation, we have $N_{\nu}\hbar\omega_{\nu}/c\sim N_e\gamma_{e0}mc$ (neglecting the momenta of the absorbed laser photons), with the electron and photon numbers $N_e$ and $N_\nu$, respectively, and the emitted photon frequency $\omega_{\nu}$. The pair production probability per laser cycle of a photon colliding with a linearly polarized laser is $P_{e^{\pm}}\sim \alpha a_0 f(\chi_{\nu})$, where  the photon quantum nonlinearity parameter is $\chi_{\nu}\approx 2a_0\hbar\omega\hbar\omega_{\nu}/(m^2c^4)$, and $f(x)$ is a fitting function  $f(x)=0.453K_{1/3}^2(4/3x)/[1+0.145x^{1/4}{\rm ln}(1+2.26x)+0.33x]$ \cite{blackburn2017scaling}. Since the radiative energy loss of electrons  increases nonlinearly with $\chi_e$, assuming electrons are reflected at the peak of the laser pulse and lose most energies there, we have $\chi_e\approx 2\gamma_{e0}a_0\hbar\omega/m$, and the typical energy of the emitted photons is $\hbar\omega_\nu\approx\gamma_{e0}m/2$ according to \cite{blackburn2017scaling}. 
Finally, we have a relation that shows only dependence on  the initial parameters: $N_{e^\pm}\sim N_\nu P_{e^{\pm}}\sim N_e\alpha a_0f(\chi_{e_0}/2)$. According to our 3D PIC simulations, we have the photon yield
\begin{eqnarray}
	N_{e^\pm}\approx 3\times10^{-3}N_ea_0f(\chi_{e_0}/2).
		\label{eq:yield}
\end{eqnarray}
The positron yield increases significantly with a higher $a_0$ and $\gamma_{e0}$, with a linear dependence with respect to $a_0f(\chi_{e_0}/2)$.  At small $a_0f(\chi_{e_0}/2)$ (small $\gamma_{e0}$), Eq. (\ref{eq:yield}) overestimates  the yield because most electrons are stopped before reaching the peak. Moreover, for the same $a_0$ and different $\gamma_{e0}$, one can obtain very good fittings for larger $\chi_{e_0}$ ($\gamma_{e0}>1000$). The indexes $m$ of the scaling $N_{e^{+}}\propto\gamma_{e0}^m$ are 2.82, 2.25, 1.91, 1.70 and 1.56 for $a_0=400$, 550, 700, 850 and 1000, respectively, in accordance with the function  $f(x)$.

\bibliography{strong_fields_bibliography}

\begin{thebibliography}{86}%
\makeatletter
\providecommand \@ifxundefined [1]{%
 \@ifx{#1\undefined}
}%
\providecommand \@ifnum [1]{%
 \ifnum #1\expandafter \@firstoftwo
 \else \expandafter \@secondoftwo
 \fi
}%
\providecommand \@ifx [1]{%
 \ifx #1\expandafter \@firstoftwo
 \else \expandafter \@secondoftwo
 \fi
}%
\providecommand \natexlab [1]{#1}%
\providecommand \enquote  [1]{``#1''}%
\providecommand \bibnamefont  [1]{#1}%
\providecommand \bibfnamefont [1]{#1}%
\providecommand \citenamefont [1]{#1}%
\providecommand \href@noop [0]{\@secondoftwo}%
\providecommand \href [0]{\begingroup \@sanitize@url \@href}%
\providecommand \@href[1]{\@@startlink{#1}\@@href}%
\providecommand \@@href[1]{\endgroup#1\@@endlink}%
\providecommand \@sanitize@url [0]{\catcode `\\12\catcode `\$12\catcode
  `\&12\catcode `\#12\catcode `\^12\catcode `\_12\catcode `\%12\relax}%
\providecommand \@@startlink[1]{}%
\providecommand \@@endlink[0]{}%
\providecommand \url  [0]{\begingroup\@sanitize@url \@url }%
\providecommand \@url [1]{\endgroup\@href {#1}{\urlprefix }}%
\providecommand \urlprefix  [0]{URL }%
\providecommand \Eprint [0]{\href }%
\providecommand \doibase [0]{https://doi.org/}%
\providecommand \selectlanguage [0]{\@gobble}%
\providecommand \bibinfo  [0]{\@secondoftwo}%
\providecommand \bibfield  [0]{\@secondoftwo}%
\providecommand \translation [1]{[#1]}%
\providecommand \BibitemOpen [0]{}%
\providecommand \bibitemStop [0]{}%
\providecommand \bibitemNoStop [0]{.\EOS\space}%
\providecommand \EOS [0]{\spacefactor3000\relax}%
\providecommand \BibitemShut  [1]{\csname bibitem#1\endcsname}%
\let\auto@bib@innerbib\@empty
\bibitem [{\citenamefont {Esarey}\ \emph {et~al.}(2009)\citenamefont {Esarey},
  \citenamefont {Schroeder},\ and\ \citenamefont {Leemans}}]{Esarey_2009}%
  \BibitemOpen
  \bibfield  {author} {\bibinfo {author} {\bibfnamefont {E.}~\bibnamefont
  {Esarey}}, \bibinfo {author} {\bibfnamefont {E.~B.}\ \bibnamefont
  {Schroeder}},\ and\ \bibinfo {author} {\bibfnamefont {W.~P.}\ \bibnamefont
  {Leemans}},\ }\bibfield  {title} {\bibinfo {title} {{Physics of laser-driven
  plasma-based electron accelerators}},\ }\href@noop {} {\bibfield  {journal}
  {\bibinfo  {journal} {Reviews of Modern Physics}\ }\textbf {\bibinfo {volume}
  {81}},\ \bibinfo {pages} {1229} (\bibinfo {year} {2009})}\BibitemShut
  {NoStop}%
\bibitem [{\citenamefont {Pukhov}\ and\ \citenamefont {Meyer-ter
  Vehn}(2002)}]{pukhov2002LWFA}%
  \BibitemOpen
  \bibfield  {author} {\bibinfo {author} {\bibfnamefont {A.}~\bibnamefont
  {Pukhov}}\ and\ \bibinfo {author} {\bibfnamefont {J.}~\bibnamefont {Meyer-ter
  Vehn}},\ }\bibfield  {title} {\bibinfo {title} {Laser wake field
  acceleration: the highly non-linear broken-wave regime},\ }\href@noop {}
  {\bibfield  {journal} {\bibinfo  {journal} {Applied Physics B}\ }\textbf
  {\bibinfo {volume} {74}},\ \bibinfo {pages} {355} (\bibinfo {year}
  {2002})}\BibitemShut {NoStop}%
\bibitem [{\citenamefont {Gordienko}\ and\ \citenamefont
  {Pukhov}(2005)}]{gordienko2005LWFA}%
  \BibitemOpen
  \bibfield  {author} {\bibinfo {author} {\bibfnamefont {S.}~\bibnamefont
  {Gordienko}}\ and\ \bibinfo {author} {\bibfnamefont {A.}~\bibnamefont
  {Pukhov}},\ }\bibfield  {title} {\bibinfo {title} {Scalings for
  ultrarelativistic laser plasmas and quasimonoenergetic electrons},\
  }\href@noop {} {\bibfield  {journal} {\bibinfo  {journal} {Physics of
  Plasmas}\ }\textbf {\bibinfo {volume} {12}} (\bibinfo {year}
  {2005})}\BibitemShut {NoStop}%
\bibitem [{\citenamefont {Lu}\ \emph {et~al.}(2007)\citenamefont {Lu},
  \citenamefont {Tzoufras}, \citenamefont {Joshi}, \citenamefont {Tsung},
  \citenamefont {Mori}, \citenamefont {Vieira}, \citenamefont {Fonseca},\ and\
  \citenamefont {Silva}}]{Lu2007}%
  \BibitemOpen
  \bibfield  {author} {\bibinfo {author} {\bibfnamefont {W.}~\bibnamefont
  {Lu}}, \bibinfo {author} {\bibfnamefont {M.}~\bibnamefont {Tzoufras}},
  \bibinfo {author} {\bibfnamefont {C.}~\bibnamefont {Joshi}}, \bibinfo
  {author} {\bibfnamefont {F.~S.}\ \bibnamefont {Tsung}}, \bibinfo {author}
  {\bibfnamefont {W.~B.}\ \bibnamefont {Mori}}, \bibinfo {author}
  {\bibfnamefont {J.}~\bibnamefont {Vieira}}, \bibinfo {author} {\bibfnamefont
  {R.~A.}\ \bibnamefont {Fonseca}},\ and\ \bibinfo {author} {\bibfnamefont
  {L.~O.}\ \bibnamefont {Silva}},\ }\bibfield  {title} {\bibinfo {title}
  {Generating multi-gev electron bunches using single stage laser wakefield
  acceleration in a 3d nonlinear regime},\ }\href@noop {} {\bibfield  {journal}
  {\bibinfo  {journal} {Phys. Rev. ST Accel. Beams}\ }\textbf {\bibinfo
  {volume} {10}},\ \bibinfo {pages} {061301} (\bibinfo {year}
  {2007})}\BibitemShut {NoStop}%
\bibitem [{\citenamefont {Steinke}\ \emph {et~al.}(2016)\citenamefont
  {Steinke}, \citenamefont {van Tilborg}, \citenamefont {Benedetti},
  \citenamefont {Geddes}, \citenamefont {Schroeder}, \citenamefont {Daniels},
  \citenamefont {Swanson}, \citenamefont {Gonsalves}, \citenamefont {Nakamura},
  \citenamefont {Matlis}, \citenamefont {Shaw}, \citenamefont {Esarey},\ and\
  \citenamefont {Leemans}}]{Steinke_2016}%
  \BibitemOpen
  \bibfield  {author} {\bibinfo {author} {\bibfnamefont {S.}~\bibnamefont
  {Steinke}}, \bibinfo {author} {\bibfnamefont {J.}~\bibnamefont {van
  Tilborg}}, \bibinfo {author} {\bibfnamefont {C.}~\bibnamefont {Benedetti}},
  \bibinfo {author} {\bibfnamefont {C.~G.~R.}\ \bibnamefont {Geddes}}, \bibinfo
  {author} {\bibfnamefont {C.~B.}\ \bibnamefont {Schroeder}}, \bibinfo {author}
  {\bibfnamefont {J.}~\bibnamefont {Daniels}}, \bibinfo {author} {\bibfnamefont
  {K.~K.}\ \bibnamefont {Swanson}}, \bibinfo {author} {\bibfnamefont {A.~J.}\
  \bibnamefont {Gonsalves}}, \bibinfo {author} {\bibfnamefont {K.}~\bibnamefont
  {Nakamura}}, \bibinfo {author} {\bibfnamefont {N.~H.}\ \bibnamefont
  {Matlis}}, \bibinfo {author} {\bibfnamefont {B.~H.}\ \bibnamefont {Shaw}},
  \bibinfo {author} {\bibfnamefont {E.}~\bibnamefont {Esarey}},\ and\ \bibinfo
  {author} {\bibfnamefont {W.~P.}\ \bibnamefont {Leemans}},\ }\bibfield
  {title} {\bibinfo {title} {{Multistage coupling of independent laser-plasma
  accelerators}},\ }\href@noop {} {\bibfield  {journal} {\bibinfo  {journal}
  {Nature}\ }\textbf {\bibinfo {volume} {530}},\ \bibinfo {pages} {190}
  (\bibinfo {year} {2016})}\BibitemShut {NoStop}%
\bibitem [{\citenamefont {Gonsalves}\ \emph {et~al.}(2019)\citenamefont
  {Gonsalves}, \citenamefont {Nakamura}, \citenamefont {Daniels}, \citenamefont
  {Benedetti}, \citenamefont {Pieronek}, \citenamefont {de~Raadt},
  \citenamefont {Steinke}, \citenamefont {Bin}, \citenamefont {Bulanov},\ and\
  \citenamefont {van Tilborg}}]{Gonsalves_2019}%
  \BibitemOpen
  \bibfield  {author} {\bibinfo {author} {\bibfnamefont {A.~J.}\ \bibnamefont
  {Gonsalves}}, \bibinfo {author} {\bibfnamefont {K.}~\bibnamefont {Nakamura}},
  \bibinfo {author} {\bibfnamefont {J.}~\bibnamefont {Daniels}}, \bibinfo
  {author} {\bibfnamefont {C.}~\bibnamefont {Benedetti}}, \bibinfo {author}
  {\bibfnamefont {C.}~\bibnamefont {Pieronek}}, \bibinfo {author}
  {\bibfnamefont {T.~C.~H.}\ \bibnamefont {de~Raadt}}, \bibinfo {author}
  {\bibfnamefont {S.}~\bibnamefont {Steinke}}, \bibinfo {author} {\bibfnamefont
  {J.~H.}\ \bibnamefont {Bin}}, \bibinfo {author} {\bibfnamefont {S.~S.}\
  \bibnamefont {Bulanov}},\ and\ \bibinfo {author} {\bibfnamefont {e.~a.}\
  \bibnamefont {van Tilborg}},\ }\bibfield  {title} {\bibinfo {title}
  {{Petawatt Laser Guiding and Electron Beam Acceleration to 8 GeV in a
  Laser-Heated Capillary Discharge Waveguide}},\ }\href@noop {} {\bibfield
  {journal} {\bibinfo  {journal} {Phys. Rev. Lett.}\ }\textbf {\bibinfo
  {volume} {122}},\ \bibinfo {pages} {084801} (\bibinfo {year}
  {2019})}\BibitemShut {NoStop}%
\bibitem [{\citenamefont {Golovanov}\ \emph {et~al.}(2023)\citenamefont
  {Golovanov}, \citenamefont {Kostyukov}, \citenamefont {Pukhov},\ and\
  \citenamefont {Malka}}]{Golovanov2023}%
  \BibitemOpen
  \bibfield  {author} {\bibinfo {author} {\bibfnamefont {A.}~\bibnamefont
  {Golovanov}}, \bibinfo {author} {\bibfnamefont {I.~Y.}\ \bibnamefont
  {Kostyukov}}, \bibinfo {author} {\bibfnamefont {A.}~\bibnamefont {Pukhov}},\
  and\ \bibinfo {author} {\bibfnamefont {V.}~\bibnamefont {Malka}},\ }\bibfield
   {title} {\bibinfo {title} {Energy-conserving theory of the blowout regime of
  plasma wakefield},\ }\href@noop {} {\bibfield  {journal} {\bibinfo  {journal}
  {Phys. Rev. Lett.}\ }\textbf {\bibinfo {volume} {130}},\ \bibinfo {pages}
  {105001} (\bibinfo {year} {2023})}\BibitemShut {NoStop}%
\bibitem [{\citenamefont {Vieira}\ and\ \citenamefont
  {Mendon\ifmmode~\mbox{\c{c}}\else \c{c}\fi{}a}(2014)}]{Vieira_2014}%
  \BibitemOpen
  \bibfield  {author} {\bibinfo {author} {\bibfnamefont {J.}~\bibnamefont
  {Vieira}}\ and\ \bibinfo {author} {\bibfnamefont {J.~T.}\ \bibnamefont
  {Mendon\ifmmode~\mbox{\c{c}}\else \c{c}\fi{}a}},\ }\bibfield  {title}
  {\bibinfo {title} {Nonlinear laser driven donut wakefields for positron and
  electron acceleration},\ }\href@noop {} {\bibfield  {journal} {\bibinfo
  {journal} {Phys. Rev. Lett.}\ }\textbf {\bibinfo {volume} {112}},\ \bibinfo
  {pages} {215001} (\bibinfo {year} {2014})}\BibitemShut {NoStop}%
\bibitem [{\citenamefont {Reichwein}\ \emph {et~al.}(2022)\citenamefont
  {Reichwein}, \citenamefont {Pukhov}, \citenamefont {Golovanov},\ and\
  \citenamefont {Kostyukov}}]{Reichwein_2022}%
  \BibitemOpen
  \bibfield  {author} {\bibinfo {author} {\bibfnamefont {L.}~\bibnamefont
  {Reichwein}}, \bibinfo {author} {\bibfnamefont {A.}~\bibnamefont {Pukhov}},
  \bibinfo {author} {\bibfnamefont {A.}~\bibnamefont {Golovanov}},\ and\
  \bibinfo {author} {\bibfnamefont {I.~Y.}\ \bibnamefont {Kostyukov}},\
  }\bibfield  {title} {\bibinfo {title} {Positron acceleration via
  laser-augmented blowouts in two-column plasma structures},\ }\href@noop {}
  {\bibfield  {journal} {\bibinfo  {journal} {Phys. Rev. E}\ }\textbf {\bibinfo
  {volume} {105}},\ \bibinfo {pages} {055207} (\bibinfo {year}
  {2022})}\BibitemShut {NoStop}%
\bibitem [{\citenamefont {Shen}\ and\ \citenamefont {Meyer-ter
  Vehn}(2001)}]{Shen2001}%
  \BibitemOpen
  \bibfield  {author} {\bibinfo {author} {\bibfnamefont {B.}~\bibnamefont
  {Shen}}\ and\ \bibinfo {author} {\bibfnamefont {J.}~\bibnamefont {Meyer-ter
  Vehn}},\ }\bibfield  {title} {\bibinfo {title} {Pair and
  \ensuremath{\gamma}-photon production from a thin foil confined by two laser
  pulses},\ }\href@noop {} {\bibfield  {journal} {\bibinfo  {journal} {Phys.
  Rev. E}\ }\textbf {\bibinfo {volume} {65}},\ \bibinfo {pages} {016405}
  (\bibinfo {year} {2001})}\BibitemShut {NoStop}%
\bibitem [{\citenamefont {Sugimoto}\ \emph {et~al.}(2023)\citenamefont
  {Sugimoto}, \citenamefont {He}, \citenamefont {Iwata}, \citenamefont {Yeh},
  \citenamefont {Tangtartharakul}, \citenamefont {Arefiev},\ and\ \citenamefont
  {Sentoku}}]{sugimoto2023positron}%
  \BibitemOpen
  \bibfield  {author} {\bibinfo {author} {\bibfnamefont {K.}~\bibnamefont
  {Sugimoto}}, \bibinfo {author} {\bibfnamefont {Y.}~\bibnamefont {He}},
  \bibinfo {author} {\bibfnamefont {N.}~\bibnamefont {Iwata}}, \bibinfo
  {author} {\bibfnamefont {I.}~\bibnamefont {Yeh}}, \bibinfo {author}
  {\bibfnamefont {K.}~\bibnamefont {Tangtartharakul}}, \bibinfo {author}
  {\bibfnamefont {A.}~\bibnamefont {Arefiev}},\ and\ \bibinfo {author}
  {\bibfnamefont {Y.}~\bibnamefont {Sentoku}},\ }\bibfield  {title} {\bibinfo
  {title} {Positron generation and acceleration in a self-organized photon
  collider enabled by an ultraintense laser pulse},\ }\href@noop {} {\bibfield
  {journal} {\bibinfo  {journal} {Physical Review Letters}\ }\textbf {\bibinfo
  {volume} {131}},\ \bibinfo {pages} {065102} (\bibinfo {year}
  {2023})}\BibitemShut {NoStop}%
\bibitem [{\citenamefont {Silva}\ \emph {et~al.}(2021)\citenamefont {Silva},
  \citenamefont {Amorim}, \citenamefont {Downer}, \citenamefont {Hogan},
  \citenamefont {Yakimenko}, \citenamefont {Zgadzaj},\ and\ \citenamefont
  {Vieira}}]{silva2021stable}%
  \BibitemOpen
  \bibfield  {author} {\bibinfo {author} {\bibfnamefont {T.}~\bibnamefont
  {Silva}}, \bibinfo {author} {\bibfnamefont {L.}~\bibnamefont {Amorim}},
  \bibinfo {author} {\bibfnamefont {M.}~\bibnamefont {Downer}}, \bibinfo
  {author} {\bibfnamefont {M.}~\bibnamefont {Hogan}}, \bibinfo {author}
  {\bibfnamefont {V.}~\bibnamefont {Yakimenko}}, \bibinfo {author}
  {\bibfnamefont {R.}~\bibnamefont {Zgadzaj}},\ and\ \bibinfo {author}
  {\bibfnamefont {J.}~\bibnamefont {Vieira}},\ }\bibfield  {title} {\bibinfo
  {title} {Stable positron acceleration in thin, warm, hollow plasma
  channels},\ }\href@noop {} {\bibfield  {journal} {\bibinfo  {journal}
  {Physical review letters}\ }\textbf {\bibinfo {volume} {127}},\ \bibinfo
  {pages} {104801} (\bibinfo {year} {2021})}\BibitemShut {NoStop}%
\bibitem [{\citenamefont {Silva}\ and\ \citenamefont
  {Vieira}(2023)}]{silva2023positron}%
  \BibitemOpen
  \bibfield  {author} {\bibinfo {author} {\bibfnamefont {T.}~\bibnamefont
  {Silva}}\ and\ \bibinfo {author} {\bibfnamefont {J.}~\bibnamefont {Vieira}},\
  }\bibfield  {title} {\bibinfo {title} {Positron acceleration in plasma waves
  driven by non-neutral fireball beams},\ }\href@noop {} {\bibfield  {journal}
  {\bibinfo  {journal} {Physical Review Accelerators and Beams}\ }\textbf
  {\bibinfo {volume} {26}},\ \bibinfo {pages} {091301} (\bibinfo {year}
  {2023})}\BibitemShut {NoStop}%
\bibitem [{\citenamefont {Xu}\ \emph {et~al.}(2020)\citenamefont {Xu},
  \citenamefont {Xiao}, \citenamefont {Lu}, \citenamefont {Hu}, \citenamefont
  {Yu}, \citenamefont {Gong}, \citenamefont {Shou}, \citenamefont {Liu},
  \citenamefont {Xie}, \citenamefont {Chen} \emph {et~al.}}]{xu2020new}%
  \BibitemOpen
  \bibfield  {author} {\bibinfo {author} {\bibfnamefont {Z.}~\bibnamefont
  {Xu}}, \bibinfo {author} {\bibfnamefont {C.}~\bibnamefont {Xiao}}, \bibinfo
  {author} {\bibfnamefont {H.}~\bibnamefont {Lu}}, \bibinfo {author}
  {\bibfnamefont {R.}~\bibnamefont {Hu}}, \bibinfo {author} {\bibfnamefont
  {J.}~\bibnamefont {Yu}}, \bibinfo {author} {\bibfnamefont {Z.}~\bibnamefont
  {Gong}}, \bibinfo {author} {\bibfnamefont {Y.}~\bibnamefont {Shou}}, \bibinfo
  {author} {\bibfnamefont {J.}~\bibnamefont {Liu}}, \bibinfo {author}
  {\bibfnamefont {C.}~\bibnamefont {Xie}}, \bibinfo {author} {\bibfnamefont
  {S.}~\bibnamefont {Chen}}, \emph {et~al.},\ }\bibfield  {title} {\bibinfo
  {title} {New injection and acceleration scheme of positrons in the
  laser-plasma bubble regime},\ }\href@noop {} {\bibfield  {journal} {\bibinfo
  {journal} {Physical Review Accelerators and Beams}\ }\textbf {\bibinfo
  {volume} {23}},\ \bibinfo {pages} {091301} (\bibinfo {year}
  {2020})}\BibitemShut {NoStop}%
\bibitem [{\citenamefont {Chen}\ \emph {et~al.}(2010)\citenamefont {Chen},
  \citenamefont {Wilks}, \citenamefont {Meyerhofer}, \citenamefont {Bonlie},
  \citenamefont {Chen}, \citenamefont {Chen}, \citenamefont {Courtois},
  \citenamefont {Elberson}, \citenamefont {Gregori}, \citenamefont {Kruer}
  \emph {et~al.}}]{chen2010relativistic}%
  \BibitemOpen
  \bibfield  {author} {\bibinfo {author} {\bibfnamefont {H.}~\bibnamefont
  {Chen}}, \bibinfo {author} {\bibfnamefont {S.}~\bibnamefont {Wilks}},
  \bibinfo {author} {\bibfnamefont {D.}~\bibnamefont {Meyerhofer}}, \bibinfo
  {author} {\bibfnamefont {J.}~\bibnamefont {Bonlie}}, \bibinfo {author}
  {\bibfnamefont {C.}~\bibnamefont {Chen}}, \bibinfo {author} {\bibfnamefont
  {S.}~\bibnamefont {Chen}}, \bibinfo {author} {\bibfnamefont {C.}~\bibnamefont
  {Courtois}}, \bibinfo {author} {\bibfnamefont {L.}~\bibnamefont {Elberson}},
  \bibinfo {author} {\bibfnamefont {G.}~\bibnamefont {Gregori}}, \bibinfo
  {author} {\bibfnamefont {W.}~\bibnamefont {Kruer}}, \emph {et~al.},\
  }\bibfield  {title} {\bibinfo {title} {Relativistic quasimonoenergetic
  positron jets from intense laser-solid interactions},\ }\href@noop {}
  {\bibfield  {journal} {\bibinfo  {journal} {Physical review letters}\
  }\textbf {\bibinfo {volume} {105}},\ \bibinfo {pages} {015003} (\bibinfo
  {year} {2010})}\BibitemShut {NoStop}%
\bibitem [{\citenamefont {Zhao}\ \emph {et~al.}(2023)\citenamefont {Zhao},
  \citenamefont {Li}, \citenamefont {Hu}, \citenamefont {Zhang}, \citenamefont
  {Cao}, \citenamefont {Sha}, \citenamefont {Shao},\ and\ \citenamefont
  {Yu}}]{zhao2023terahertz}%
  \BibitemOpen
  \bibfield  {author} {\bibinfo {author} {\bibfnamefont {J.}~\bibnamefont
  {Zhao}}, \bibinfo {author} {\bibfnamefont {Q.-N.}\ \bibnamefont {Li}},
  \bibinfo {author} {\bibfnamefont {Y.-T.}\ \bibnamefont {Hu}}, \bibinfo
  {author} {\bibfnamefont {H.}~\bibnamefont {Zhang}}, \bibinfo {author}
  {\bibfnamefont {Y.}~\bibnamefont {Cao}}, \bibinfo {author} {\bibfnamefont
  {R.}~\bibnamefont {Sha}}, \bibinfo {author} {\bibfnamefont {F.-Q.}\
  \bibnamefont {Shao}},\ and\ \bibinfo {author} {\bibfnamefont {T.-P.}\
  \bibnamefont {Yu}},\ }\bibfield  {title} {\bibinfo {title} {Terahertz-driven
  positron acceleration assisted by ultra-intense lasers},\ }\href@noop {}
  {\bibfield  {journal} {\bibinfo  {journal} {Optics Express}\ }\textbf
  {\bibinfo {volume} {31}},\ \bibinfo {pages} {23171} (\bibinfo {year}
  {2023})}\BibitemShut {NoStop}%
\bibitem [{\citenamefont {Martinez}\ \emph {et~al.}(2023)\citenamefont
  {Martinez}, \citenamefont {Barbosa},\ and\ \citenamefont
  {Vranic}}]{martinez2023creation}%
  \BibitemOpen
  \bibfield  {author} {\bibinfo {author} {\bibfnamefont {B.}~\bibnamefont
  {Martinez}}, \bibinfo {author} {\bibfnamefont {B.}~\bibnamefont {Barbosa}},\
  and\ \bibinfo {author} {\bibfnamefont {M.}~\bibnamefont {Vranic}},\
  }\bibfield  {title} {\bibinfo {title} {Creation and direct laser acceleration
  of positrons in a single stage},\ }\href@noop {} {\bibfield  {journal}
  {\bibinfo  {journal} {Physical Review Accelerators and Beams}\ }\textbf
  {\bibinfo {volume} {26}},\ \bibinfo {pages} {011301} (\bibinfo {year}
  {2023})}\BibitemShut {NoStop}%
\bibitem [{\citenamefont {Gamiz}\ \emph {et~al.}(2024)\citenamefont {Gamiz},
  \citenamefont {Babjak}, \citenamefont {Martinez},\ and\ \citenamefont
  {Vrani{\'c}}}]{gamiz2024improved}%
  \BibitemOpen
  \bibfield  {author} {\bibinfo {author} {\bibfnamefont {L.~I.~I.}\
  \bibnamefont {Gamiz}}, \bibinfo {author} {\bibfnamefont {R.}~\bibnamefont
  {Babjak}}, \bibinfo {author} {\bibfnamefont {B.}~\bibnamefont {Martinez}},\
  and\ \bibinfo {author} {\bibfnamefont {M.}~\bibnamefont {Vrani{\'c}}},\
  }\bibfield  {title} {\bibinfo {title} {{Improved Bethe-Heitler positron
  creation and retention by combining direct laser acceleration and solid
  target interaction within a gas jet}},\ }\href@noop {} {\bibfield  {journal}
  {\bibinfo  {journal} {arXiv preprint arXiv:2411.17455}\ } (\bibinfo {year}
  {2024})}\BibitemShut {NoStop}%
\bibitem [{\citenamefont {Esarey}\ \emph {et~al.}(1995)\citenamefont {Esarey},
  \citenamefont {Sprangle},\ and\ \citenamefont {Krall}}]{Esarey_1995}%
  \BibitemOpen
  \bibfield  {author} {\bibinfo {author} {\bibfnamefont {E.}~\bibnamefont
  {Esarey}}, \bibinfo {author} {\bibfnamefont {P.}~\bibnamefont {Sprangle}},\
  and\ \bibinfo {author} {\bibfnamefont {J.}~\bibnamefont {Krall}},\ }\bibfield
   {title} {\bibinfo {title} {{Laser acceleration of electrons in vacuum}},\
  }\href@noop {} {\bibfield  {journal} {\bibinfo  {journal} {Physical Review
  E}\ }\textbf {\bibinfo {volume} {52}},\ \bibinfo {pages} {5443} (\bibinfo
  {year} {1995})}\BibitemShut {NoStop}%
\bibitem [{\citenamefont {Salamin}\ and\ \citenamefont
  {Keitel}(2002)}]{Salamin_2002}%
  \BibitemOpen
  \bibfield  {author} {\bibinfo {author} {\bibfnamefont {Y.~I.}\ \bibnamefont
  {Salamin}}\ and\ \bibinfo {author} {\bibfnamefont {C.~H.}\ \bibnamefont
  {Keitel}},\ }\bibfield  {title} {\bibinfo {title} {Electron acceleration by a
  tightly focused laser beam},\ }\href@noop {} {\bibfield  {journal} {\bibinfo
  {journal} {Phys. Rev. Lett.}\ }\textbf {\bibinfo {volume} {88}},\ \bibinfo
  {pages} {095005} (\bibinfo {year} {2002})}\BibitemShut {NoStop}%
\bibitem [{\citenamefont {Marceau}\ \emph {et~al.}(2012)\citenamefont
  {Marceau}, \citenamefont {April},\ and\ \citenamefont
  {Pich{\'{e}}}}]{Marceau_2012}%
  \BibitemOpen
  \bibfield  {author} {\bibinfo {author} {\bibfnamefont {V.}~\bibnamefont
  {Marceau}}, \bibinfo {author} {\bibfnamefont {A.}~\bibnamefont {April}},\
  and\ \bibinfo {author} {\bibfnamefont {M.}~\bibnamefont {Pich{\'{e}}}},\
  }\bibfield  {title} {\bibinfo {title} {{Electron acceleration driven by
  ultrashort and nonparaxial radially polarized laser pulses}},\ }\href@noop {}
  {\bibfield  {journal} {\bibinfo  {journal} {Optics Letters}\ }\textbf
  {\bibinfo {volume} {37}},\ \bibinfo {pages} {2442} (\bibinfo {year}
  {2012})}\BibitemShut {NoStop}%
\bibitem [{\citenamefont {Powell}\ \emph {et~al.}(2024)\citenamefont {Powell},
  \citenamefont {Jolly}, \citenamefont {Valli{\`{e}}res}, \citenamefont
  {Fillion-Gourdeau}, \citenamefont {Payeur}, \citenamefont {Fourmaux},
  \citenamefont {Lytova}, \citenamefont {Pich{\'{e}}}, \citenamefont {Ibrahim},
  \citenamefont {MacLean},\ and\ \citenamefont
  {L{\'{e}}gar{\'{e}}}}]{Powell_2024}%
  \BibitemOpen
  \bibfield  {author} {\bibinfo {author} {\bibfnamefont {J.}~\bibnamefont
  {Powell}}, \bibinfo {author} {\bibfnamefont {S.~W.}\ \bibnamefont {Jolly}},
  \bibinfo {author} {\bibfnamefont {S.}~\bibnamefont {Valli{\`{e}}res}},
  \bibinfo {author} {\bibfnamefont {F.}~\bibnamefont {Fillion-Gourdeau}},
  \bibinfo {author} {\bibfnamefont {S.}~\bibnamefont {Payeur}}, \bibinfo
  {author} {\bibfnamefont {S.}~\bibnamefont {Fourmaux}}, \bibinfo {author}
  {\bibfnamefont {M.}~\bibnamefont {Lytova}}, \bibinfo {author} {\bibfnamefont
  {M.}~\bibnamefont {Pich{\'{e}}}}, \bibinfo {author} {\bibfnamefont
  {H.}~\bibnamefont {Ibrahim}}, \bibinfo {author} {\bibfnamefont
  {S.}~\bibnamefont {MacLean}},\ and\ \bibinfo {author} {\bibfnamefont
  {F.}~\bibnamefont {L{\'{e}}gar{\'{e}}}},\ }\bibfield  {title} {\bibinfo
  {title} {{Relativistic Electrons from Vacuum Laser Acceleration Using Tightly
  Focused Radially Polarized Beams}},\ }\href@noop {} {\bibfield  {journal}
  {\bibinfo  {journal} {Physical Review Letters}\ }\textbf {\bibinfo {volume}
  {133}},\ \bibinfo {pages} {155001} (\bibinfo {year} {2024})}\BibitemShut
  {NoStop}%
\bibitem [{\citenamefont {Th{\'{e}}venet}\ \emph {et~al.}(2016)\citenamefont
  {Th{\'{e}}venet}, \citenamefont {Leblanc}, \citenamefont {Kahaly},
  \citenamefont {Vincenti}, \citenamefont {Vernier}, \citenamefont
  {Qu{\'{e}}r{\'{e}}},\ and\ \citenamefont {Faure}}]{Thevenet_2015}%
  \BibitemOpen
  \bibfield  {author} {\bibinfo {author} {\bibfnamefont {M.}~\bibnamefont
  {Th{\'{e}}venet}}, \bibinfo {author} {\bibfnamefont {A.}~\bibnamefont
  {Leblanc}}, \bibinfo {author} {\bibfnamefont {S.}~\bibnamefont {Kahaly}},
  \bibinfo {author} {\bibfnamefont {H.}~\bibnamefont {Vincenti}}, \bibinfo
  {author} {\bibfnamefont {A.}~\bibnamefont {Vernier}}, \bibinfo {author}
  {\bibfnamefont {F.}~\bibnamefont {Qu{\'{e}}r{\'{e}}}},\ and\ \bibinfo
  {author} {\bibfnamefont {J.}~\bibnamefont {Faure}},\ }\bibfield  {title}
  {\bibinfo {title} {{Vacuum laser acceleration of relativistic electrons using
  plasma mirror injectors}},\ }\href@noop {} {\bibfield  {journal} {\bibinfo
  {journal} {Nature Physics}\ }\textbf {\bibinfo {volume} {12}},\ \bibinfo
  {pages} {355} (\bibinfo {year} {2016})}\BibitemShut {NoStop}%
\bibitem [{\citenamefont {Singh}\ \emph {et~al.}(2022)\citenamefont {Singh},
  \citenamefont {Li}, \citenamefont {Huang}, \citenamefont {Moreau},
  \citenamefont {Hollinger}, \citenamefont {Junghans}, \citenamefont {Favalli},
  \citenamefont {Calvi}, \citenamefont {Wang}, \citenamefont {Wang},
  \citenamefont {Song}, \citenamefont {Rocca}, \citenamefont {Reinovsky},\ and\
  \citenamefont {Palaniyappan}}]{Singh_2022}%
  \BibitemOpen
  \bibfield  {author} {\bibinfo {author} {\bibfnamefont {P.~K.}\ \bibnamefont
  {Singh}}, \bibinfo {author} {\bibfnamefont {F.~Y.}\ \bibnamefont {Li}},
  \bibinfo {author} {\bibfnamefont {C.~K.}\ \bibnamefont {Huang}}, \bibinfo
  {author} {\bibfnamefont {A.}~\bibnamefont {Moreau}}, \bibinfo {author}
  {\bibfnamefont {R.}~\bibnamefont {Hollinger}}, \bibinfo {author}
  {\bibfnamefont {A.}~\bibnamefont {Junghans}}, \bibinfo {author}
  {\bibfnamefont {A.}~\bibnamefont {Favalli}}, \bibinfo {author} {\bibfnamefont
  {C.}~\bibnamefont {Calvi}}, \bibinfo {author} {\bibfnamefont
  {S.}~\bibnamefont {Wang}}, \bibinfo {author} {\bibfnamefont {Y.}~\bibnamefont
  {Wang}}, \bibinfo {author} {\bibfnamefont {H.}~\bibnamefont {Song}}, \bibinfo
  {author} {\bibfnamefont {J.~J.}\ \bibnamefont {Rocca}}, \bibinfo {author}
  {\bibfnamefont {R.~E.}\ \bibnamefont {Reinovsky}},\ and\ \bibinfo {author}
  {\bibfnamefont {S.}~\bibnamefont {Palaniyappan}},\ }\bibfield  {title}
  {\bibinfo {title} {{Vacuum laser acceleration of super-ponderomotive
  electrons using relativistic transparency injection}},\ }\href@noop {}
  {\bibfield  {journal} {\bibinfo  {journal} {Nature Communications}\ }\textbf
  {\bibinfo {volume} {13}},\ \bibinfo {pages} {1} (\bibinfo {year}
  {2022})}\BibitemShut {NoStop}%
\bibitem [{\citenamefont {{De Andres}}\ \emph {et~al.}(2024)\citenamefont {{De
  Andres}}, \citenamefont {Bhadoria}, \citenamefont {Marmolejo}, \citenamefont
  {Muschet}, \citenamefont {Fischer}, \citenamefont {{Reza Barzegar}},
  \citenamefont {Blackburn}, \citenamefont {Gonoskov}, \citenamefont
  {Hanstorp}, \citenamefont {Marklund},\ and\ \citenamefont
  {Veisz}}]{DeAndres_2024}%
  \BibitemOpen
  \bibfield  {author} {\bibinfo {author} {\bibfnamefont {A.}~\bibnamefont {{De
  Andres}}}, \bibinfo {author} {\bibfnamefont {S.}~\bibnamefont {Bhadoria}},
  \bibinfo {author} {\bibfnamefont {J.~T.}\ \bibnamefont {Marmolejo}}, \bibinfo
  {author} {\bibfnamefont {A.}~\bibnamefont {Muschet}}, \bibinfo {author}
  {\bibfnamefont {P.}~\bibnamefont {Fischer}}, \bibinfo {author} {\bibfnamefont
  {H.}~\bibnamefont {{Reza Barzegar}}}, \bibinfo {author} {\bibfnamefont
  {T.}~\bibnamefont {Blackburn}}, \bibinfo {author} {\bibfnamefont
  {A.}~\bibnamefont {Gonoskov}}, \bibinfo {author} {\bibfnamefont
  {D.}~\bibnamefont {Hanstorp}}, \bibinfo {author} {\bibfnamefont
  {M.}~\bibnamefont {Marklund}},\ and\ \bibinfo {author} {\bibfnamefont
  {L.}~\bibnamefont {Veisz}},\ }\bibfield  {title} {\bibinfo {title}
  {{Unforeseen advantage of looser focusing in vacuum laser acceleration}},\
  }\href@noop {} {\bibfield  {journal} {\bibinfo  {journal} {Communications
  Physics}\ }\textbf {\bibinfo {volume} {7}},\ \bibinfo {pages} {1} (\bibinfo
  {year} {2024})}\BibitemShut {NoStop}%
\bibitem [{\citenamefont {Bucksbaum}\ \emph {et~al.}(1987)\citenamefont
  {Bucksbaum}, \citenamefont {Bashkansky},\ and\ \citenamefont
  {McIlrath}}]{Bucksbaum_1987}%
  \BibitemOpen
  \bibfield  {author} {\bibinfo {author} {\bibfnamefont {P.~H.}\ \bibnamefont
  {Bucksbaum}}, \bibinfo {author} {\bibfnamefont {M.}~\bibnamefont
  {Bashkansky}},\ and\ \bibinfo {author} {\bibfnamefont {T.~J.}\ \bibnamefont
  {McIlrath}},\ }\bibfield  {title} {\bibinfo {title} {{Scattering of Electrons
  by Intense Coherent Light}},\ }\href@noop {} {\bibfield  {journal} {\bibinfo
  {journal} {Phys. Rev. Lett.}\ }\textbf {\bibinfo {volume} {58}},\ \bibinfo
  {pages} {349} (\bibinfo {year} {1987})}\BibitemShut {NoStop}%
\bibitem [{\citenamefont {Hartemann}\ \emph {et~al.}(1995)\citenamefont
  {Hartemann}, \citenamefont {Fochs}, \citenamefont {{Le Sage}}, \citenamefont
  {Luhmann}, \citenamefont {Woodworth}, \citenamefont {Perry}, \citenamefont
  {Chen},\ and\ \citenamefont {Kerman}}]{Hartemann_1995}%
  \BibitemOpen
  \bibfield  {author} {\bibinfo {author} {\bibfnamefont {F.~V.}\ \bibnamefont
  {Hartemann}}, \bibinfo {author} {\bibfnamefont {S.~N.}\ \bibnamefont
  {Fochs}}, \bibinfo {author} {\bibfnamefont {G.~P.}\ \bibnamefont {{Le
  Sage}}}, \bibinfo {author} {\bibfnamefont {N.~C.}\ \bibnamefont {Luhmann}},
  \bibinfo {author} {\bibfnamefont {J.~G.}\ \bibnamefont {Woodworth}}, \bibinfo
  {author} {\bibfnamefont {M.~D.}\ \bibnamefont {Perry}}, \bibinfo {author}
  {\bibfnamefont {Y.~J.}\ \bibnamefont {Chen}},\ and\ \bibinfo {author}
  {\bibfnamefont {A.~K.}\ \bibnamefont {Kerman}},\ }\bibfield  {title}
  {\bibinfo {title} {{Nonlinear ponderomotive scattering of relativistic
  electrons by an intense laser field at focus}},\ }\href@noop {} {\bibfield
  {journal} {\bibinfo  {journal} {Physical Review E}\ }\textbf {\bibinfo
  {volume} {51}},\ \bibinfo {pages} {4833} (\bibinfo {year}
  {1995})}\BibitemShut {NoStop}%
\bibitem [{\citenamefont {Malka}\ \emph {et~al.}(1997)\citenamefont {Malka},
  \citenamefont {Lefebvre},\ and\ \citenamefont {Miquel}}]{Malka_1997}%
  \BibitemOpen
  \bibfield  {author} {\bibinfo {author} {\bibfnamefont {V.}~\bibnamefont
  {Malka}}, \bibinfo {author} {\bibfnamefont {E.}~\bibnamefont {Lefebvre}},\
  and\ \bibinfo {author} {\bibfnamefont {J.~L.}\ \bibnamefont {Miquel}},\
  }\bibfield  {title} {\bibinfo {title} {{Experimental Observation of Electrons
  Accelerated in Vacuum to Relativistic Energies by a High-Intensity Laser}},\
  }\href@noop {} {\bibfield  {journal} {\bibinfo  {journal} {Physical Review
  Letters}\ }\textbf {\bibinfo {volume} {78}},\ \bibinfo {pages} {3314}
  (\bibinfo {year} {1997})}\BibitemShut {NoStop}%
\bibitem [{\citenamefont {Mora}\ and\ \citenamefont
  {Quesnel}(1998)}]{Mora_1998}%
  \BibitemOpen
  \bibfield  {author} {\bibinfo {author} {\bibfnamefont {P.}~\bibnamefont
  {Mora}}\ and\ \bibinfo {author} {\bibfnamefont {B.}~\bibnamefont {Quesnel}},\
  }\bibfield  {title} {\bibinfo {title} {Comment on ``experimental observation
  of electrons accelerated in vacuum to relativistic energies by a
  high-intensity laser''},\ }\href@noop {} {\bibfield  {journal} {\bibinfo
  {journal} {Phys. Rev. Lett.}\ }\textbf {\bibinfo {volume} {80}},\ \bibinfo
  {pages} {1351} (\bibinfo {year} {1998})}\BibitemShut {NoStop}%
\bibitem [{\citenamefont {Quesnel}\ and\ \citenamefont
  {Mora}(1998)}]{Quesnel_1998}%
  \BibitemOpen
  \bibfield  {author} {\bibinfo {author} {\bibfnamefont {B.}~\bibnamefont
  {Quesnel}}\ and\ \bibinfo {author} {\bibfnamefont {P.}~\bibnamefont {Mora}},\
  }\bibfield  {title} {\bibinfo {title} {{Theory and simulation of the
  interaction of ultraintense laser pulses with electrons in vacuum}},\
  }\href@noop {} {\bibfield  {journal} {\bibinfo  {journal} {Physical Review
  E}\ }\textbf {\bibinfo {volume} {58}},\ \bibinfo {pages} {3719} (\bibinfo
  {year} {1998})}\BibitemShut {NoStop}%
\bibitem [{\citenamefont {Stupakov}\ and\ \citenamefont
  {Zolotorev}(2001)}]{Stupakov_2001}%
  \BibitemOpen
  \bibfield  {author} {\bibinfo {author} {\bibfnamefont {G.~V.}\ \bibnamefont
  {Stupakov}}\ and\ \bibinfo {author} {\bibfnamefont {M.~S.}\ \bibnamefont
  {Zolotorev}},\ }\bibfield  {title} {\bibinfo {title} {{Ponderomotive laser
  acceleration and focusing in vacuum for generation of attosecond electron
  bunches}},\ }\href@noop {} {\bibfield  {journal} {\bibinfo  {journal}
  {Physical review letters}\ }\textbf {\bibinfo {volume} {86}},\ \bibinfo
  {pages} {5274} (\bibinfo {year} {2001})}\BibitemShut {NoStop}%
\bibitem [{\citenamefont {He}\ \emph {et~al.}(2003)\citenamefont {He},
  \citenamefont {Yu}, \citenamefont {Lu}, \citenamefont {Xu}, \citenamefont
  {Qian}, \citenamefont {Shen}, \citenamefont {Yuan}, \citenamefont {Li},\ and\
  \citenamefont {Xu}}]{He_2003}%
  \BibitemOpen
  \bibfield  {author} {\bibinfo {author} {\bibfnamefont {F.}~\bibnamefont
  {He}}, \bibinfo {author} {\bibfnamefont {W.}~\bibnamefont {Yu}}, \bibinfo
  {author} {\bibfnamefont {P.}~\bibnamefont {Lu}}, \bibinfo {author}
  {\bibfnamefont {H.}~\bibnamefont {Xu}}, \bibinfo {author} {\bibfnamefont
  {L.}~\bibnamefont {Qian}}, \bibinfo {author} {\bibfnamefont {B.}~\bibnamefont
  {Shen}}, \bibinfo {author} {\bibfnamefont {X.}~\bibnamefont {Yuan}}, \bibinfo
  {author} {\bibfnamefont {R.}~\bibnamefont {Li}},\ and\ \bibinfo {author}
  {\bibfnamefont {Z.}~\bibnamefont {Xu}},\ }\bibfield  {title} {\bibinfo
  {title} {Ponderomotive acceleration of electrons by a tightly focused intense
  laser beam},\ }\href@noop {} {\bibfield  {journal} {\bibinfo  {journal}
  {Phys. Rev. E}\ }\textbf {\bibinfo {volume} {68}},\ \bibinfo {pages} {046407}
  (\bibinfo {year} {2003})}\BibitemShut {NoStop}%
\bibitem [{\citenamefont {Moore}\ \emph {et~al.}(1999)\citenamefont {Moore},
  \citenamefont {Ting}, \citenamefont {McNaught}, \citenamefont {Qiu},
  \citenamefont {Burris},\ and\ \citenamefont {Sprangle}}]{Moore_1999}%
  \BibitemOpen
  \bibfield  {author} {\bibinfo {author} {\bibfnamefont {C.~I.}\ \bibnamefont
  {Moore}}, \bibinfo {author} {\bibfnamefont {A.}~\bibnamefont {Ting}},
  \bibinfo {author} {\bibfnamefont {S.~J.}\ \bibnamefont {McNaught}}, \bibinfo
  {author} {\bibfnamefont {J.}~\bibnamefont {Qiu}}, \bibinfo {author}
  {\bibfnamefont {H.~R.}\ \bibnamefont {Burris}},\ and\ \bibinfo {author}
  {\bibfnamefont {P.}~\bibnamefont {Sprangle}},\ }\bibfield  {title} {\bibinfo
  {title} {{A Laser-Accelerator Injector Based on Laser Ionization and
  Ponderomotive Acceleration of Electrons}},\ }\href@noop {} {\bibfield
  {journal} {\bibinfo  {journal} {Physical Review Letters}\ }\textbf {\bibinfo
  {volume} {82}},\ \bibinfo {pages} {1688} (\bibinfo {year}
  {1999})}\BibitemShut {NoStop}%
\bibitem [{\citenamefont {Hu}\ and\ \citenamefont {Starace}(2002)}]{Hu_2002}%
  \BibitemOpen
  \bibfield  {author} {\bibinfo {author} {\bibfnamefont {S.~X.}\ \bibnamefont
  {Hu}}\ and\ \bibinfo {author} {\bibfnamefont {A.~F.}\ \bibnamefont
  {Starace}},\ }\bibfield  {title} {\bibinfo {title} {{GeV Electrons from
  Ultraintense Laser Interaction with Highly Charged Ions}},\ }\href@noop {}
  {\bibfield  {journal} {\bibinfo  {journal} {Physical Review Letters}\
  }\textbf {\bibinfo {volume} {88}},\ \bibinfo {pages} {245003} (\bibinfo
  {year} {2002})}\BibitemShut {NoStop}%
\bibitem [{\citenamefont {{Di Piazza}}\ \emph {et~al.}(2012)\citenamefont {{Di
  Piazza}}, \citenamefont {M\"uller}, \citenamefont {Hatsagortsyan},\ and\
  \citenamefont {Keitel}}]{RMP_2012}%
  \BibitemOpen
  \bibfield  {author} {\bibinfo {author} {\bibfnamefont {A.}~\bibnamefont {{Di
  Piazza}}}, \bibinfo {author} {\bibfnamefont {C.}~\bibnamefont {M\"uller}},
  \bibinfo {author} {\bibfnamefont {K.~Z.}\ \bibnamefont {Hatsagortsyan}},\
  and\ \bibinfo {author} {\bibfnamefont {C.~H.}\ \bibnamefont {Keitel}},\
  }\bibfield  {title} {\bibinfo {title} {Extremely high-intensity laser
  interactions with fundamental quantum systems},\ }\href@noop {} {\bibfield
  {journal} {\bibinfo  {journal} {Rev. Mod. Phys.}\ }\textbf {\bibinfo {volume}
  {84}},\ \bibinfo {pages} {1177} (\bibinfo {year} {2012})}\BibitemShut
  {NoStop}%
\bibitem [{\citenamefont {Dodin}\ and\ \citenamefont
  {Fisch}(2003)}]{Dodin_2003}%
  \BibitemOpen
  \bibfield  {author} {\bibinfo {author} {\bibfnamefont {I.~Y.}\ \bibnamefont
  {Dodin}}\ and\ \bibinfo {author} {\bibfnamefont {N.~J.}\ \bibnamefont
  {Fisch}},\ }\bibfield  {title} {\bibinfo {title} {{Relativistic electron
  acceleration in focused laser fields after above-threshold ionization.}},\
  }\href@noop {} {\bibfield  {journal} {\bibinfo  {journal} {Physical review.
  E}\ }\textbf {\bibinfo {volume} {68}},\ \bibinfo {pages} {056402} (\bibinfo
  {year} {2003})}\BibitemShut {NoStop}%
\bibitem [{\citenamefont {Maltsev}\ and\ \citenamefont
  {Ditmire}(2003)}]{Maltsev_2003}%
  \BibitemOpen
  \bibfield  {author} {\bibinfo {author} {\bibfnamefont {A.}~\bibnamefont
  {Maltsev}}\ and\ \bibinfo {author} {\bibfnamefont {T.}~\bibnamefont
  {Ditmire}},\ }\bibfield  {title} {\bibinfo {title} {{Above Threshold
  Ionization in Tightly Focused, Strongly Relativistic Laser Fields}},\
  }\href@noop {} {\bibfield  {journal} {\bibinfo  {journal} {Physical review
  letters}\ }\textbf {\bibinfo {volume} {90}},\ \bibinfo {pages} {053002}
  (\bibinfo {year} {2003})}\BibitemShut {NoStop}%
\bibitem [{\citenamefont {Gordon}\ \emph {et~al.}(2017)\citenamefont {Gordon},
  \citenamefont {Palastro},\ and\ \citenamefont {Hafizi}}]{Gordon_2017}%
  \BibitemOpen
  \bibfield  {author} {\bibinfo {author} {\bibfnamefont {D.~F.}\ \bibnamefont
  {Gordon}}, \bibinfo {author} {\bibfnamefont {J.~P.}\ \bibnamefont
  {Palastro}},\ and\ \bibinfo {author} {\bibfnamefont {B.}~\bibnamefont
  {Hafizi}},\ }\bibfield  {title} {\bibinfo {title} {{Superponderomotive regime
  of tunneling ionization}},\ }\href@noop {} {\bibfield  {journal} {\bibinfo
  {journal} {Physical Review A}\ }\textbf {\bibinfo {volume} {95}},\ \bibinfo
  {pages} {033403} (\bibinfo {year} {2017})}\BibitemShut {NoStop}%
\bibitem [{\citenamefont {Yandow}\ \emph {et~al.}(2019)\citenamefont {Yandow},
  \citenamefont {Toncian},\ and\ \citenamefont {Ditmire}}]{Yandow_2019}%
  \BibitemOpen
  \bibfield  {author} {\bibinfo {author} {\bibfnamefont {A.}~\bibnamefont
  {Yandow}}, \bibinfo {author} {\bibfnamefont {T.}~\bibnamefont {Toncian}},\
  and\ \bibinfo {author} {\bibfnamefont {T.}~\bibnamefont {Ditmire}},\
  }\bibfield  {title} {\bibinfo {title} {{Direct laser ion acceleration and
  above-threshold ionization at intensities from $10^{21}$ W/cm$^2$ to $3\times
  10^{23}$ W/cm$^2$3}},\ }\href@noop {} {\bibfield  {journal} {\bibinfo
  {journal} {Physical Review A}\ }\textbf {\bibinfo {volume} {100}},\ \bibinfo
  {pages} {053406} (\bibinfo {year} {2019})}\BibitemShut {NoStop}%
\bibitem [{\citenamefont {Yandow}\ \emph {et~al.}(2024)\citenamefont {Yandow},
  \citenamefont {Ha}, \citenamefont {Aniculaesei}, \citenamefont {Smith},
  \citenamefont {Richmond}, \citenamefont {Spinks}, \citenamefont {Quevedo},
  \citenamefont {Bruce}, \citenamefont {Darilek}, \citenamefont {Chang},
  \citenamefont {Garcia}, \citenamefont {Gaul}, \citenamefont {Donovan},
  \citenamefont {Hegelich},\ and\ \citenamefont {Ditmire}}]{Yandow_2024}%
  \BibitemOpen
  \bibfield  {author} {\bibinfo {author} {\bibfnamefont {A.}~\bibnamefont
  {Yandow}}, \bibinfo {author} {\bibfnamefont {T.~N.}\ \bibnamefont {Ha}},
  \bibinfo {author} {\bibfnamefont {C.}~\bibnamefont {Aniculaesei}}, \bibinfo
  {author} {\bibfnamefont {H.~L.}\ \bibnamefont {Smith}}, \bibinfo {author}
  {\bibfnamefont {C.~G.}\ \bibnamefont {Richmond}}, \bibinfo {author}
  {\bibfnamefont {M.~M.}\ \bibnamefont {Spinks}}, \bibinfo {author}
  {\bibfnamefont {H.~J.}\ \bibnamefont {Quevedo}}, \bibinfo {author}
  {\bibfnamefont {S.}~\bibnamefont {Bruce}}, \bibinfo {author} {\bibfnamefont
  {M.}~\bibnamefont {Darilek}}, \bibinfo {author} {\bibfnamefont
  {C.}~\bibnamefont {Chang}}, \bibinfo {author} {\bibfnamefont {D.~A.}\
  \bibnamefont {Garcia}}, \bibinfo {author} {\bibfnamefont {E.}~\bibnamefont
  {Gaul}}, \bibinfo {author} {\bibfnamefont {M.~E.}\ \bibnamefont {Donovan}},
  \bibinfo {author} {\bibfnamefont {B.~M.}\ \bibnamefont {Hegelich}},\ and\
  \bibinfo {author} {\bibfnamefont {T.}~\bibnamefont {Ditmire}},\ }\bibfield
  {title} {\bibinfo {title} {Above-threshold ionization at laser intensity
  greater than ${10}^{20} \mathrm{W}/{\mathrm{cm}}^{2}$},\ }\href@noop {}
  {\bibfield  {journal} {\bibinfo  {journal} {Phys. Rev. A}\ }\textbf {\bibinfo
  {volume} {109}},\ \bibinfo {pages} {023119} (\bibinfo {year}
  {2024})}\BibitemShut {NoStop}%
\bibitem [{\citenamefont {Yoon}\ \emph {et~al.}(2021)\citenamefont {Yoon},
  \citenamefont {Kim}, \citenamefont {Choi}, \citenamefont {Sung},
  \citenamefont {Lee}, \citenamefont {Lee},\ and\ \citenamefont
  {Nam}}]{Yoon_2021}%
  \BibitemOpen
  \bibfield  {author} {\bibinfo {author} {\bibfnamefont {J.~W.}\ \bibnamefont
  {Yoon}}, \bibinfo {author} {\bibfnamefont {Y.~G.}\ \bibnamefont {Kim}},
  \bibinfo {author} {\bibfnamefont {I.~W.}\ \bibnamefont {Choi}}, \bibinfo
  {author} {\bibfnamefont {J.~H.}\ \bibnamefont {Sung}}, \bibinfo {author}
  {\bibfnamefont {H.~W.}\ \bibnamefont {Lee}}, \bibinfo {author} {\bibfnamefont
  {S.~K.}\ \bibnamefont {Lee}},\ and\ \bibinfo {author} {\bibfnamefont {C.~H.}\
  \bibnamefont {Nam}},\ }\bibfield  {title} {\bibinfo {title} {{Realization of
  laser intensity over 10$^{23}$ W/cm$^2$}},\ }\href@noop {} {\bibfield
  {journal} {\bibinfo  {journal} {Optica}\ }\textbf {\bibinfo {volume} {8}},\
  \bibinfo {pages} {630} (\bibinfo {year} {2021})}\BibitemShut {NoStop}%
\bibitem [{\citenamefont {Fedotov}\ \emph {et~al.}(2014)\citenamefont
  {Fedotov}, \citenamefont {Elkina}, \citenamefont {Gelfer}, \citenamefont
  {Narozhny},\ and\ \citenamefont {Ruhl}}]{Fedotov_2014}%
  \BibitemOpen
  \bibfield  {author} {\bibinfo {author} {\bibfnamefont {A.~M.}\ \bibnamefont
  {Fedotov}}, \bibinfo {author} {\bibfnamefont {N.~V.}\ \bibnamefont {Elkina}},
  \bibinfo {author} {\bibfnamefont {E.~G.}\ \bibnamefont {Gelfer}}, \bibinfo
  {author} {\bibfnamefont {N.~B.}\ \bibnamefont {Narozhny}},\ and\ \bibinfo
  {author} {\bibfnamefont {H.}~\bibnamefont {Ruhl}},\ }\bibfield  {title}
  {\bibinfo {title} {Radiation friction versus ponderomotive effect},\
  }\href@noop {} {\bibfield  {journal} {\bibinfo  {journal} {Phys. Rev. A}\
  }\textbf {\bibinfo {volume} {90}},\ \bibinfo {pages} {053847} (\bibinfo
  {year} {2014})}\BibitemShut {NoStop}%
\bibitem [{\citenamefont {Poder}\ \emph {et~al.}(2018)\citenamefont {Poder},
  \citenamefont {Tamburini}, \citenamefont {Sarri}, \citenamefont {Di~Piazza},
  \citenamefont {Kuschel}, \citenamefont {Baird}, \citenamefont {Behm},
  \citenamefont {Bohlen}, \citenamefont {Cole}, \citenamefont {Corvan},
  \citenamefont {Duff}, \citenamefont {Gerstmayr}, \citenamefont {Keitel},
  \citenamefont {Krushelnick}, \citenamefont {Mangles}, \citenamefont
  {McKenna}, \citenamefont {Murphy}, \citenamefont {Najmudin}, \citenamefont
  {Ridgers}, \citenamefont {Samarin}, \citenamefont {Symes}, \citenamefont
  {Thomas}, \citenamefont {Warwick},\ and\ \citenamefont {Zepf}}]{Poder_2018}%
  \BibitemOpen
  \bibfield  {author} {\bibinfo {author} {\bibfnamefont {K.}~\bibnamefont
  {Poder}}, \bibinfo {author} {\bibfnamefont {M.}~\bibnamefont {Tamburini}},
  \bibinfo {author} {\bibfnamefont {G.}~\bibnamefont {Sarri}}, \bibinfo
  {author} {\bibfnamefont {A.}~\bibnamefont {Di~Piazza}}, \bibinfo {author}
  {\bibfnamefont {S.}~\bibnamefont {Kuschel}}, \bibinfo {author} {\bibfnamefont
  {C.~D.}\ \bibnamefont {Baird}}, \bibinfo {author} {\bibfnamefont
  {K.}~\bibnamefont {Behm}}, \bibinfo {author} {\bibfnamefont {S.}~\bibnamefont
  {Bohlen}}, \bibinfo {author} {\bibfnamefont {J.~M.}\ \bibnamefont {Cole}},
  \bibinfo {author} {\bibfnamefont {D.~J.}\ \bibnamefont {Corvan}}, \bibinfo
  {author} {\bibfnamefont {M.}~\bibnamefont {Duff}}, \bibinfo {author}
  {\bibfnamefont {E.}~\bibnamefont {Gerstmayr}}, \bibinfo {author}
  {\bibfnamefont {C.~H.}\ \bibnamefont {Keitel}}, \bibinfo {author}
  {\bibfnamefont {K.}~\bibnamefont {Krushelnick}}, \bibinfo {author}
  {\bibfnamefont {S.~P.~D.}\ \bibnamefont {Mangles}}, \bibinfo {author}
  {\bibfnamefont {P.}~\bibnamefont {McKenna}}, \bibinfo {author} {\bibfnamefont
  {C.~D.}\ \bibnamefont {Murphy}}, \bibinfo {author} {\bibfnamefont
  {Z.}~\bibnamefont {Najmudin}}, \bibinfo {author} {\bibfnamefont {C.~P.}\
  \bibnamefont {Ridgers}}, \bibinfo {author} {\bibfnamefont {G.~M.}\
  \bibnamefont {Samarin}}, \bibinfo {author} {\bibfnamefont {D.~R.}\
  \bibnamefont {Symes}}, \bibinfo {author} {\bibfnamefont {A.~G.~R.}\
  \bibnamefont {Thomas}}, \bibinfo {author} {\bibfnamefont {J.}~\bibnamefont
  {Warwick}},\ and\ \bibinfo {author} {\bibfnamefont {M.}~\bibnamefont
  {Zepf}},\ }\bibfield  {title} {\bibinfo {title} {Experimental signatures of
  the quantum nature of radiation reaction in the field of an ultraintense
  laser},\ }\href@noop {} {\bibfield  {journal} {\bibinfo  {journal} {Phys.
  Rev. X}\ }\textbf {\bibinfo {volume} {8}},\ \bibinfo {pages} {031004}
  (\bibinfo {year} {2018})}\BibitemShut {NoStop}%
\bibitem [{\citenamefont {Cole}\ \emph {et~al.}(2018)\citenamefont {Cole},
  \citenamefont {Behm}, \citenamefont {Gerstmayr}, \citenamefont {Blackburn},
  \citenamefont {Wood}, \citenamefont {Baird}, \citenamefont {Duff},
  \citenamefont {Harvey}, \citenamefont {Ilderton}, \citenamefont {Joglekar},
  \citenamefont {Krushelnick}, \citenamefont {Kuschel}, \citenamefont
  {Marklund}, \citenamefont {McKenna}, \citenamefont {Murphy}, \citenamefont
  {Poder}, \citenamefont {Ridgers}, \citenamefont {Samarin}, \citenamefont
  {Sarri}, \citenamefont {Symes}, \citenamefont {Thomas}, \citenamefont
  {Warwick}, \citenamefont {Zepf}, \citenamefont {Najmudin},\ and\
  \citenamefont {Mangles}}]{Cole_2018}%
  \BibitemOpen
  \bibfield  {author} {\bibinfo {author} {\bibfnamefont {J.~M.}\ \bibnamefont
  {Cole}}, \bibinfo {author} {\bibfnamefont {K.~T.}\ \bibnamefont {Behm}},
  \bibinfo {author} {\bibfnamefont {E.}~\bibnamefont {Gerstmayr}}, \bibinfo
  {author} {\bibfnamefont {T.~G.}\ \bibnamefont {Blackburn}}, \bibinfo {author}
  {\bibfnamefont {J.~C.}\ \bibnamefont {Wood}}, \bibinfo {author}
  {\bibfnamefont {C.~D.}\ \bibnamefont {Baird}}, \bibinfo {author}
  {\bibfnamefont {M.~J.}\ \bibnamefont {Duff}}, \bibinfo {author}
  {\bibfnamefont {C.}~\bibnamefont {Harvey}}, \bibinfo {author} {\bibfnamefont
  {A.}~\bibnamefont {Ilderton}}, \bibinfo {author} {\bibfnamefont {A.~S.}\
  \bibnamefont {Joglekar}}, \bibinfo {author} {\bibfnamefont {K.}~\bibnamefont
  {Krushelnick}}, \bibinfo {author} {\bibfnamefont {S.}~\bibnamefont
  {Kuschel}}, \bibinfo {author} {\bibfnamefont {M.}~\bibnamefont {Marklund}},
  \bibinfo {author} {\bibfnamefont {P.}~\bibnamefont {McKenna}}, \bibinfo
  {author} {\bibfnamefont {C.~D.}\ \bibnamefont {Murphy}}, \bibinfo {author}
  {\bibfnamefont {K.}~\bibnamefont {Poder}}, \bibinfo {author} {\bibfnamefont
  {C.~P.}\ \bibnamefont {Ridgers}}, \bibinfo {author} {\bibfnamefont {G.~M.}\
  \bibnamefont {Samarin}}, \bibinfo {author} {\bibfnamefont {G.}~\bibnamefont
  {Sarri}}, \bibinfo {author} {\bibfnamefont {D.~R.}\ \bibnamefont {Symes}},
  \bibinfo {author} {\bibfnamefont {A.~G.~R.}\ \bibnamefont {Thomas}}, \bibinfo
  {author} {\bibfnamefont {J.}~\bibnamefont {Warwick}}, \bibinfo {author}
  {\bibfnamefont {M.}~\bibnamefont {Zepf}}, \bibinfo {author} {\bibfnamefont
  {Z.}~\bibnamefont {Najmudin}},\ and\ \bibinfo {author} {\bibfnamefont
  {S.~P.~D.}\ \bibnamefont {Mangles}},\ }\bibfield  {title} {\bibinfo {title}
  {Experimental evidence of radiation reaction in the collision of a
  high-intensity laser pulse with a laser-wakefield accelerated electron
  beam},\ }\href@noop {} {\bibfield  {journal} {\bibinfo  {journal} {Phys. Rev.
  X}\ }\textbf {\bibinfo {volume} {8}},\ \bibinfo {pages} {011020} (\bibinfo
  {year} {2018})}\BibitemShut {NoStop}%
\bibitem [{\citenamefont {Ritus}(1985)}]{Ritus_1985}%
  \BibitemOpen
  \bibfield  {author} {\bibinfo {author} {\bibfnamefont {V.~I.}\ \bibnamefont
  {Ritus}},\ }\bibfield  {title} {\bibinfo {title} {Quantum effects of the
  interaction of elementary particles with an inense electromagnetic field},\
  }\href@noop {} {\bibfield  {journal} {\bibinfo  {journal} {J. Sov. Laser
  Res.}\ }\textbf {\bibinfo {volume} {6}},\ \bibinfo {pages} {497} (\bibinfo
  {year} {1985})}\BibitemShut {NoStop}%
\bibitem [{\citenamefont {Burke}\ \emph {et~al.}(1997)\citenamefont {Burke},
  \citenamefont {Field}, \citenamefont {Horton-Smith}, \citenamefont {Spencer},
  \citenamefont {Walz}, \citenamefont {Berridge}, \citenamefont {Bugg},
  \citenamefont {Shmakov}, \citenamefont {Weidemann}, \citenamefont {Bula},
  \citenamefont {McDonald}, \citenamefont {Prebys}, \citenamefont {Bamber},
  \citenamefont {Boege}, \citenamefont {Koffas}, \citenamefont {Kotseroglou},
  \citenamefont {Melissinos}, \citenamefont {Meyerhofer}, \citenamefont
  {Reis},\ and\ \citenamefont {Ragg}}]{Burke_1997}%
  \BibitemOpen
  \bibfield  {author} {\bibinfo {author} {\bibfnamefont {D.~L.}\ \bibnamefont
  {Burke}}, \bibinfo {author} {\bibfnamefont {R.~C.}\ \bibnamefont {Field}},
  \bibinfo {author} {\bibfnamefont {G.}~\bibnamefont {Horton-Smith}}, \bibinfo
  {author} {\bibfnamefont {J.~E.}\ \bibnamefont {Spencer}}, \bibinfo {author}
  {\bibfnamefont {D.}~\bibnamefont {Walz}}, \bibinfo {author} {\bibfnamefont
  {S.~C.}\ \bibnamefont {Berridge}}, \bibinfo {author} {\bibfnamefont {W.~M.}\
  \bibnamefont {Bugg}}, \bibinfo {author} {\bibfnamefont {K.}~\bibnamefont
  {Shmakov}}, \bibinfo {author} {\bibfnamefont {A.~W.}\ \bibnamefont
  {Weidemann}}, \bibinfo {author} {\bibfnamefont {C.}~\bibnamefont {Bula}},
  \bibinfo {author} {\bibfnamefont {K.~T.}\ \bibnamefont {McDonald}}, \bibinfo
  {author} {\bibfnamefont {E.~J.}\ \bibnamefont {Prebys}}, \bibinfo {author}
  {\bibfnamefont {C.}~\bibnamefont {Bamber}}, \bibinfo {author} {\bibfnamefont
  {S.~J.}\ \bibnamefont {Boege}}, \bibinfo {author} {\bibfnamefont
  {T.}~\bibnamefont {Koffas}}, \bibinfo {author} {\bibfnamefont
  {T.}~\bibnamefont {Kotseroglou}}, \bibinfo {author} {\bibfnamefont {A.~C.}\
  \bibnamefont {Melissinos}}, \bibinfo {author} {\bibfnamefont {D.~D.}\
  \bibnamefont {Meyerhofer}}, \bibinfo {author} {\bibfnamefont {D.~A.}\
  \bibnamefont {Reis}},\ and\ \bibinfo {author} {\bibfnamefont
  {W.}~\bibnamefont {Ragg}},\ }\bibfield  {title} {\bibinfo {title} {Positron
  production in multiphoton light-by-light scattering},\ }\href@noop {}
  {\bibfield  {journal} {\bibinfo  {journal} {Phys. Rev. Lett.}\ }\textbf
  {\bibinfo {volume} {79}},\ \bibinfo {pages} {1626} (\bibinfo {year}
  {1997})}\BibitemShut {NoStop}%
\bibitem [{\citenamefont {Bu}\ \emph {et~al.}(2021)\citenamefont {Bu},
  \citenamefont {Ji}, \citenamefont {Lei}, \citenamefont {Hu}, \citenamefont
  {Zhang},\ and\ \citenamefont {Shen}}]{Bu_2021}%
  \BibitemOpen
  \bibfield  {author} {\bibinfo {author} {\bibfnamefont {Z.}~\bibnamefont
  {Bu}}, \bibinfo {author} {\bibfnamefont {L.}~\bibnamefont {Ji}}, \bibinfo
  {author} {\bibfnamefont {S.}~\bibnamefont {Lei}}, \bibinfo {author}
  {\bibfnamefont {H.}~\bibnamefont {Hu}}, \bibinfo {author} {\bibfnamefont
  {X.}~\bibnamefont {Zhang}},\ and\ \bibinfo {author} {\bibfnamefont
  {B.}~\bibnamefont {Shen}},\ }\bibfield  {title} {\bibinfo {title} {{Twisted
  Breit-Wheeler electron-positron pair creation via vortex gamma photons}},\
  }\href@noop {} {\bibfield  {journal} {\bibinfo  {journal} {Phys. Rev. Res.}\
  }\textbf {\bibinfo {volume} {3}},\ \bibinfo {pages} {043159} (\bibinfo {year}
  {2021})}\BibitemShut {NoStop}%
\bibitem [{\citenamefont {Eckey}\ \emph {et~al.}(2024)\citenamefont {Eckey},
  \citenamefont {Golub}, \citenamefont {Salgado}, \citenamefont
  {Villalba-Ch\'avez}, \citenamefont {Voitkiv}, \citenamefont {Zepf},\ and\
  \citenamefont {M\"uller}}]{Eckey_2024}%
  \BibitemOpen
  \bibfield  {author} {\bibinfo {author} {\bibfnamefont {A.}~\bibnamefont
  {Eckey}}, \bibinfo {author} {\bibfnamefont {A.}~\bibnamefont {Golub}},
  \bibinfo {author} {\bibfnamefont {F.~C.}\ \bibnamefont {Salgado}}, \bibinfo
  {author} {\bibfnamefont {S.}~\bibnamefont {Villalba-Ch\'avez}}, \bibinfo
  {author} {\bibfnamefont {A.~B.}\ \bibnamefont {Voitkiv}}, \bibinfo {author}
  {\bibfnamefont {M.}~\bibnamefont {Zepf}},\ and\ \bibinfo {author}
  {\bibfnamefont {C.}~\bibnamefont {M\"uller}},\ }\bibfield  {title} {\bibinfo
  {title} {{Impact of laser focusing and radiation reaction on particle spectra
  from nonlinear Breit-Wheeler pair production in the nonperturbative
  regime}},\ }\href@noop {} {\bibfield  {journal} {\bibinfo  {journal} {Phys.
  Rev. A}\ }\textbf {\bibinfo {volume} {110}},\ \bibinfo {pages} {043113}
  (\bibinfo {year} {2024})}\BibitemShut {NoStop}%
\bibitem [{\citenamefont {{Di~Piazza}}\ \emph {et~al.}(2009)\citenamefont
  {{Di~Piazza}}, \citenamefont {Hatsagortsyan},\ and\ \citenamefont
  {Keitel}}]{DiPiazza_2009}%
  \BibitemOpen
  \bibfield  {author} {\bibinfo {author} {\bibfnamefont {A.}~\bibnamefont
  {{Di~Piazza}}}, \bibinfo {author} {\bibfnamefont {K.~Z.}\ \bibnamefont
  {Hatsagortsyan}},\ and\ \bibinfo {author} {\bibfnamefont {C.~H.}\
  \bibnamefont {Keitel}},\ }\bibfield  {title} {\bibinfo {title} {Strong
  signatures of radiation reaction below the radiation-dominated regime},\
  }\href@noop {} {\bibfield  {journal} {\bibinfo  {journal} {Phys. Rev. Lett.}\
  }\textbf {\bibinfo {volume} {102}},\ \bibinfo {pages} {254802} (\bibinfo
  {year} {2009})}\BibitemShut {NoStop}%
\bibitem [{\citenamefont {Li}\ \emph {et~al.}(2015)\citenamefont {Li},
  \citenamefont {Hatsagortsyan}, \citenamefont {Galow},\ and\ \citenamefont
  {Keitel}}]{Li_2015}%
  \BibitemOpen
  \bibfield  {author} {\bibinfo {author} {\bibfnamefont {J.-X.}\ \bibnamefont
  {Li}}, \bibinfo {author} {\bibfnamefont {K.~Z.}\ \bibnamefont
  {Hatsagortsyan}}, \bibinfo {author} {\bibfnamefont {B.~J.}\ \bibnamefont
  {Galow}},\ and\ \bibinfo {author} {\bibfnamefont {C.~H.}\ \bibnamefont
  {Keitel}},\ }\bibfield  {title} {\bibinfo {title} {Attosecond gamma-ray
  pulses via nonlinear compton scattering in the radiation-dominated regime},\
  }\href@noop {} {\bibfield  {journal} {\bibinfo  {journal} {Phys. Rev. Lett.}\
  }\textbf {\bibinfo {volume} {115}},\ \bibinfo {pages} {204801} (\bibinfo
  {year} {2015})}\BibitemShut {NoStop}%
\bibitem [{\citenamefont {Li}\ \emph {et~al.}(2018)\citenamefont {Li},
  \citenamefont {Chen}, \citenamefont {Hatsagortsyan},\ and\ \citenamefont
  {Keitel}}]{Li_2018}%
  \BibitemOpen
  \bibfield  {author} {\bibinfo {author} {\bibfnamefont {J.-X.}\ \bibnamefont
  {Li}}, \bibinfo {author} {\bibfnamefont {Y.-Y.}\ \bibnamefont {Chen}},
  \bibinfo {author} {\bibfnamefont {K.~Z.}\ \bibnamefont {Hatsagortsyan}},\
  and\ \bibinfo {author} {\bibfnamefont {C.~H.}\ \bibnamefont {Keitel}},\
  }\bibfield  {title} {\bibinfo {title} {Single-shot carrier-envelope phase
  determination of long superintense laser pulses},\ }\href@noop {} {\bibfield
  {journal} {\bibinfo  {journal} {Phys. Rev. Lett.}\ }\textbf {\bibinfo
  {volume} {120}},\ \bibinfo {pages} {124803} (\bibinfo {year}
  {2018})}\BibitemShut {NoStop}%
\bibitem [{\citenamefont {Zhuang}\ \emph {et~al.}(2023)\citenamefont {Zhuang},
  \citenamefont {Chen}, \citenamefont {Li}, \citenamefont {Hatsagortsyan},\
  and\ \citenamefont {Keitel}}]{Zhuang_2023}%
  \BibitemOpen
  \bibfield  {author} {\bibinfo {author} {\bibfnamefont {K.-H.}\ \bibnamefont
  {Zhuang}}, \bibinfo {author} {\bibfnamefont {Y.-Y.}\ \bibnamefont {Chen}},
  \bibinfo {author} {\bibfnamefont {Y.-F.}\ \bibnamefont {Li}}, \bibinfo
  {author} {\bibfnamefont {K.~Z.}\ \bibnamefont {Hatsagortsyan}},\ and\
  \bibinfo {author} {\bibfnamefont {C.~H.}\ \bibnamefont {Keitel}},\ }\bibfield
   {title} {\bibinfo {title} {Laser-driven lepton polarization in the quantum
  radiation-dominated reflection regime},\ }\href@noop {} {\bibfield  {journal}
  {\bibinfo  {journal} {Phys. Rev. D}\ }\textbf {\bibinfo {volume} {108}},\
  \bibinfo {pages} {033001} (\bibinfo {year} {2023})}\BibitemShut {NoStop}%
\bibitem [{\citenamefont {Zhang}(2023)}]{Zhang_2023}%
  \BibitemOpen
  \bibfield  {author} {\bibinfo {author} {\bibfnamefont {B.}~\bibnamefont
  {Zhang}},\ }\bibfield  {title} {\bibinfo {title} {The physics of fast radio
  bursts},\ }\href@noop {} {\bibfield  {journal} {\bibinfo  {journal} {Rev.
  Mod. Phys.}\ }\textbf {\bibinfo {volume} {95}},\ \bibinfo {pages} {035005}
  (\bibinfo {year} {2023})}\BibitemShut {NoStop}%
\bibitem [{\citenamefont {Piran}(2005)}]{Piran_2005}%
  \BibitemOpen
  \bibfield  {author} {\bibinfo {author} {\bibfnamefont {T.}~\bibnamefont
  {Piran}},\ }\bibfield  {title} {\bibinfo {title} {The physics of gamma-ray
  bursts},\ }\href@noop {} {\bibfield  {journal} {\bibinfo  {journal} {Rev.
  Mod. Phys.}\ }\textbf {\bibinfo {volume} {76}},\ \bibinfo {pages} {1143}
  (\bibinfo {year} {2005})}\BibitemShut {NoStop}%
\bibitem [{\citenamefont {Letessier-Selvon}\ and\ \citenamefont
  {Stanev}(2011)}]{Letessier_2011}%
  \BibitemOpen
  \bibfield  {author} {\bibinfo {author} {\bibfnamefont {A.}~\bibnamefont
  {Letessier-Selvon}}\ and\ \bibinfo {author} {\bibfnamefont {T.}~\bibnamefont
  {Stanev}},\ }\bibfield  {title} {\bibinfo {title} {Ultrahigh energy cosmic
  rays},\ }\href@noop {} {\bibfield  {journal} {\bibinfo  {journal} {Rev. Mod.
  Phys.}\ }\textbf {\bibinfo {volume} {83}},\ \bibinfo {pages} {907} (\bibinfo
  {year} {2011})}\BibitemShut {NoStop}%
\bibitem [{\citenamefont {Vay}\ \emph {et~al.}(2018)\citenamefont {Vay},
  \citenamefont {Almgren}, \citenamefont {Bell}, \citenamefont {Ge},
  \citenamefont {Grote}, \citenamefont {Hogan}, \citenamefont {Kononenko},
  \citenamefont {Lehe}, \citenamefont {Myers}, \citenamefont {Ng},
  \citenamefont {Park}, \citenamefont {Ryne}, \citenamefont {Shapoval},
  \citenamefont {Thévenet},\ and\ \citenamefont {Zhang}}]{Warp-X}%
  \BibitemOpen
  \bibfield  {author} {\bibinfo {author} {\bibfnamefont {J.-L.}\ \bibnamefont
  {Vay}}, \bibinfo {author} {\bibfnamefont {A.}~\bibnamefont {Almgren}},
  \bibinfo {author} {\bibfnamefont {J.}~\bibnamefont {Bell}}, \bibinfo {author}
  {\bibfnamefont {L.}~\bibnamefont {Ge}}, \bibinfo {author} {\bibfnamefont
  {D.}~\bibnamefont {Grote}}, \bibinfo {author} {\bibfnamefont
  {M.}~\bibnamefont {Hogan}}, \bibinfo {author} {\bibfnamefont
  {O.}~\bibnamefont {Kononenko}}, \bibinfo {author} {\bibfnamefont
  {R.}~\bibnamefont {Lehe}}, \bibinfo {author} {\bibfnamefont {A.}~\bibnamefont
  {Myers}}, \bibinfo {author} {\bibfnamefont {C.}~\bibnamefont {Ng}}, \bibinfo
  {author} {\bibfnamefont {J.}~\bibnamefont {Park}}, \bibinfo {author}
  {\bibfnamefont {R.}~\bibnamefont {Ryne}}, \bibinfo {author} {\bibfnamefont
  {O.}~\bibnamefont {Shapoval}}, \bibinfo {author} {\bibfnamefont
  {M.}~\bibnamefont {Thévenet}},\ and\ \bibinfo {author} {\bibfnamefont
  {W.}~\bibnamefont {Zhang}},\ }\bibfield  {title} {\bibinfo {title} {{Warp-X:
  A new exascale computing platform for beam–plasma simulations}},\
  }\href@noop {} {\bibfield  {journal} {\bibinfo  {journal} {Nuclear
  Instruments and Methods in Physics Research Section A}\ }\textbf {\bibinfo
  {volume} {909}},\ \bibinfo {pages} {476} (\bibinfo {year}
  {2018})}\BibitemShut {NoStop}%
\bibitem [{\citenamefont {Cowan}\ \emph {et~al.}(2013)\citenamefont {Cowan},
  \citenamefont {Bruhwiler}, \citenamefont {Cary}, \citenamefont
  {Cormier-Michel},\ and\ \citenamefont {Geddes}}]{Cowan_2013}%
  \BibitemOpen
  \bibfield  {author} {\bibinfo {author} {\bibfnamefont {B.~M.}\ \bibnamefont
  {Cowan}}, \bibinfo {author} {\bibfnamefont {D.~L.}\ \bibnamefont
  {Bruhwiler}}, \bibinfo {author} {\bibfnamefont {J.~R.}\ \bibnamefont {Cary}},
  \bibinfo {author} {\bibfnamefont {E.}~\bibnamefont {Cormier-Michel}},\ and\
  \bibinfo {author} {\bibfnamefont {C.~G.~R.}\ \bibnamefont {Geddes}},\
  }\bibfield  {title} {\bibinfo {title} {Generalized algorithm for control of
  numerical dispersion in explicit time-domain electromagnetic simulations},\
  }\href@noop {} {\bibfield  {journal} {\bibinfo  {journal} {Phys. Rev. ST
  Accel. Beams}\ }\textbf {\bibinfo {volume} {16}},\ \bibinfo {pages} {041303}
  (\bibinfo {year} {2013})}\BibitemShut {NoStop}%
\bibitem [{Sol()}]{Solver}%
  \BibitemOpen
  \href@noop {} {}\bibinfo {howpublished} {{To improve accuracy and suppress
  numerical Cherenkov radiation, we employed the modified finite-difference
  Maxwell-equations solver CKC, and the pseudo-spectral Maxwell equations
  solver PSATD, which gave comparable results. Numerical convergence has been
  checked by doubling the resolution in both longitudinal and transverse
  directions. We found that both the initial self fields and the effect of beam
  charge and current on evolving fields can be neglected because of the
  dominance of the laser field.}}\BibitemShut {Stop}%
\bibitem [{\citenamefont {Cordes}\ and\ \citenamefont
  {Chatterjee}(2019)}]{Cordes_2019}%
  \BibitemOpen
  \bibfield  {author} {\bibinfo {author} {\bibfnamefont {J.~M.}\ \bibnamefont
  {Cordes}}\ and\ \bibinfo {author} {\bibfnamefont {S.}~\bibnamefont
  {Chatterjee}},\ }\bibfield  {title} {\bibinfo {title} {Fast radio bursts: an
  extragalactic enigma},\ }\href@noop {} {\bibfield  {journal} {\bibinfo
  {journal} {Annual Review of Astronomy and Astrophysics}\ }\textbf {\bibinfo
  {volume} {57}},\ \bibinfo {pages} {417} (\bibinfo {year} {2019})}\BibitemShut
  {NoStop}%
\bibitem [{\citenamefont {Zhang}(2020)}]{Zhang_2020}%
  \BibitemOpen
  \bibfield  {author} {\bibinfo {author} {\bibfnamefont {B.}~\bibnamefont
  {Zhang}},\ }\bibfield  {title} {\bibinfo {title} {The physical mechanisms of
  fast radio bursts},\ }\href@noop {} {\bibfield  {journal} {\bibinfo
  {journal} {Nature}\ }\textbf {\bibinfo {volume} {587}},\ \bibinfo {pages}
  {45} (\bibinfo {year} {2020})}\BibitemShut {NoStop}%
\bibitem [{\citenamefont {Lyne}\ and\ \citenamefont
  {Graham-Smith}(2012)}]{lyne2012pulsar}%
  \BibitemOpen
  \bibfield  {author} {\bibinfo {author} {\bibfnamefont {A.}~\bibnamefont
  {Lyne}}\ and\ \bibinfo {author} {\bibfnamefont {F.}~\bibnamefont
  {Graham-Smith}},\ }\href@noop {} {\emph {\bibinfo {title} {Pulsar
  astronomy}}},\ \bibinfo {number} {48}\ (\bibinfo  {publisher} {Cambridge
  University Press},\ \bibinfo {year} {2012})\BibitemShut {NoStop}%
\bibitem [{\citenamefont {Zhang}(2018)}]{zhang2018FRB}%
  \BibitemOpen
  \bibfield  {author} {\bibinfo {author} {\bibfnamefont {B.}~\bibnamefont
  {Zhang}},\ }\bibfield  {title} {\bibinfo {title} {Fast radio burst energetics
  and detectability from high redshifts},\ }\href@noop {} {\bibfield  {journal}
  {\bibinfo  {journal} {The Astrophysical Journal Letters}\ }\textbf {\bibinfo
  {volume} {867}},\ \bibinfo {pages} {L21} (\bibinfo {year}
  {2018})}\BibitemShut {NoStop}%
\bibitem [{\citenamefont {Nimmo}\ \emph {et~al.}(2022)\citenamefont {Nimmo},
  \citenamefont {Hessels}, \citenamefont {Kirsten}, \citenamefont {Keimpema},
  \citenamefont {Cordes}, \citenamefont {Snelders}, \citenamefont {Hewitt},
  \citenamefont {Karuppusamy}, \citenamefont {Archibald}, \citenamefont
  {Bezrukovs} \emph {et~al.}}]{nimmo2022FRB}%
  \BibitemOpen
  \bibfield  {author} {\bibinfo {author} {\bibfnamefont {K.}~\bibnamefont
  {Nimmo}}, \bibinfo {author} {\bibfnamefont {J.}~\bibnamefont {Hessels}},
  \bibinfo {author} {\bibfnamefont {F.}~\bibnamefont {Kirsten}}, \bibinfo
  {author} {\bibfnamefont {A.}~\bibnamefont {Keimpema}}, \bibinfo {author}
  {\bibfnamefont {J.}~\bibnamefont {Cordes}}, \bibinfo {author} {\bibfnamefont
  {M.}~\bibnamefont {Snelders}}, \bibinfo {author} {\bibfnamefont
  {D.}~\bibnamefont {Hewitt}}, \bibinfo {author} {\bibfnamefont
  {R.}~\bibnamefont {Karuppusamy}}, \bibinfo {author} {\bibfnamefont
  {A.}~\bibnamefont {Archibald}}, \bibinfo {author} {\bibfnamefont
  {V.}~\bibnamefont {Bezrukovs}}, \emph {et~al.},\ }\bibfield  {title}
  {\bibinfo {title} {Burst timescales and luminosities as links between young
  pulsars and fast radio bursts},\ }\href@noop {} {\bibfield  {journal}
  {\bibinfo  {journal} {Nature Astronomy}\ }\textbf {\bibinfo {volume} {6}},\
  \bibinfo {pages} {393} (\bibinfo {year} {2022})}\BibitemShut {NoStop}%
\bibitem [{\citenamefont {Kumar}\ \emph {et~al.}(2017)\citenamefont {Kumar},
  \citenamefont {Lu},\ and\ \citenamefont {Bhattacharya}}]{kumar2017FRB}%
  \BibitemOpen
  \bibfield  {author} {\bibinfo {author} {\bibfnamefont {P.}~\bibnamefont
  {Kumar}}, \bibinfo {author} {\bibfnamefont {W.}~\bibnamefont {Lu}},\ and\
  \bibinfo {author} {\bibfnamefont {M.}~\bibnamefont {Bhattacharya}},\
  }\bibfield  {title} {\bibinfo {title} {Fast radio burst source properties and
  curvature radiation model},\ }\href@noop {} {\bibfield  {journal} {\bibinfo
  {journal} {Monthly Notices of the Royal Astronomical Society}\ }\textbf
  {\bibinfo {volume} {468}},\ \bibinfo {pages} {2726} (\bibinfo {year}
  {2017})}\BibitemShut {NoStop}%
\bibitem [{\citenamefont {Gajjar}\ \emph {et~al.}(2018)\citenamefont {Gajjar},
  \citenamefont {Siemion}, \citenamefont {Price}, \citenamefont {Law},
  \citenamefont {Michilli}, \citenamefont {Hessels}, \citenamefont
  {Chatterjee}, \citenamefont {Archibald}, \citenamefont {Bower}, \citenamefont
  {Brinkman} \emph {et~al.}}]{gajjar2018FRB_linear}%
  \BibitemOpen
  \bibfield  {author} {\bibinfo {author} {\bibfnamefont {V.}~\bibnamefont
  {Gajjar}}, \bibinfo {author} {\bibfnamefont {A.}~\bibnamefont {Siemion}},
  \bibinfo {author} {\bibfnamefont {D.}~\bibnamefont {Price}}, \bibinfo
  {author} {\bibfnamefont {C.}~\bibnamefont {Law}}, \bibinfo {author}
  {\bibfnamefont {D.}~\bibnamefont {Michilli}}, \bibinfo {author}
  {\bibfnamefont {J.}~\bibnamefont {Hessels}}, \bibinfo {author} {\bibfnamefont
  {S.}~\bibnamefont {Chatterjee}}, \bibinfo {author} {\bibfnamefont
  {A.}~\bibnamefont {Archibald}}, \bibinfo {author} {\bibfnamefont
  {G.}~\bibnamefont {Bower}}, \bibinfo {author} {\bibfnamefont
  {C.}~\bibnamefont {Brinkman}}, \emph {et~al.},\ }\bibfield  {title} {\bibinfo
  {title} {Highest frequency detection of frb 121102 at 4--8 ghz using the
  breakthrough listen digital backend at the green bank telescope},\
  }\href@noop {} {\bibfield  {journal} {\bibinfo  {journal} {The Astrophysical
  Journal}\ }\textbf {\bibinfo {volume} {863}},\ \bibinfo {pages} {2} (\bibinfo
  {year} {2018})}\BibitemShut {NoStop}%
\bibitem [{\citenamefont {Lyubarsky}(2020)}]{lyubarsky2020FRB_reconnection}%
  \BibitemOpen
  \bibfield  {author} {\bibinfo {author} {\bibfnamefont {Y.}~\bibnamefont
  {Lyubarsky}},\ }\bibfield  {title} {\bibinfo {title} {Fast radio bursts from
  reconnection in a magnetar magnetosphere},\ }\href@noop {} {\bibfield
  {journal} {\bibinfo  {journal} {The Astrophysical Journal}\ }\textbf
  {\bibinfo {volume} {897}},\ \bibinfo {pages} {1} (\bibinfo {year}
  {2020})}\BibitemShut {NoStop}%
\bibitem [{\citenamefont {Popov}\ \emph {et~al.}(2017)\citenamefont {Popov},
  \citenamefont {Rudnitskii},\ and\ \citenamefont {Soglasnov}}]{popov2017GP}%
  \BibitemOpen
  \bibfield  {author} {\bibinfo {author} {\bibfnamefont {M.}~\bibnamefont
  {Popov}}, \bibinfo {author} {\bibfnamefont {A.}~\bibnamefont {Rudnitskii}},\
  and\ \bibinfo {author} {\bibfnamefont {V.}~\bibnamefont {Soglasnov}},\
  }\bibfield  {title} {\bibinfo {title} {Giant pulses of the crab nebula pulsar
  as an indicator of a strong electromagnetic wave},\ }\href@noop {} {\bibfield
   {journal} {\bibinfo  {journal} {Astronomy Reports}\ }\textbf {\bibinfo
  {volume} {61}},\ \bibinfo {pages} {178} (\bibinfo {year} {2017})}\BibitemShut
  {NoStop}%
\bibitem [{\citenamefont {Yang}\ and\ \citenamefont
  {Zhang}(2020)}]{yang2020FRB}%
  \BibitemOpen
  \bibfield  {author} {\bibinfo {author} {\bibfnamefont {Y.-P.}\ \bibnamefont
  {Yang}}\ and\ \bibinfo {author} {\bibfnamefont {B.}~\bibnamefont {Zhang}},\
  }\bibfield  {title} {\bibinfo {title} {Fast radio bursts as strong waves
  interacting with the ambient medium},\ }\href@noop {} {\bibfield  {journal}
  {\bibinfo  {journal} {The Astrophysical Journal Letters}\ }\textbf {\bibinfo
  {volume} {892}},\ \bibinfo {pages} {L10} (\bibinfo {year}
  {2020})}\BibitemShut {NoStop}%
\bibitem [{\citenamefont {Lyutikov}(2021)}]{Lyutikov_2021}%
  \BibitemOpen
  \bibfield  {author} {\bibinfo {author} {\bibfnamefont {M.}~\bibnamefont
  {Lyutikov}},\ }\bibfield  {title} {\bibinfo {title} {{Magnetic loading of
  magnetars’ flares}},\ }\href@noop {} {\bibfield  {journal} {\bibinfo
  {journal} {Monthly Notices of the Royal Astronomical Society}\ }\textbf
  {\bibinfo {volume} {509}},\ \bibinfo {pages} {2689} (\bibinfo {year}
  {2021})}\BibitemShut {NoStop}%
\bibitem [{\citenamefont {{Goldreich}}\ and\ \citenamefont
  {{Julian}}(1969)}]{Goldreich_1969}%
  \BibitemOpen
  \bibfield  {author} {\bibinfo {author} {\bibfnamefont {P.}~\bibnamefont
  {{Goldreich}}}\ and\ \bibinfo {author} {\bibfnamefont {W.~H.}\ \bibnamefont
  {{Julian}}},\ }\bibfield  {title} {\bibinfo {title} {{Pulsar
  Electrodynamics}},\ }\href@noop {} {\bibfield  {journal} {\bibinfo  {journal}
  {\apj}\ }\textbf {\bibinfo {volume} {157}},\ \bibinfo {pages} {869} (\bibinfo
  {year} {1969})}\BibitemShut {NoStop}%
\bibitem [{\citenamefont {Katz}(1982)}]{katz1982magnetar}%
  \BibitemOpen
  \bibfield  {author} {\bibinfo {author} {\bibfnamefont {J.}~\bibnamefont
  {Katz}},\ }\bibfield  {title} {\bibinfo {title} {Physical processes in
  gamma-ray bursts},\ }\href@noop {} {\bibfield  {journal} {\bibinfo  {journal}
  {Astrophysical Journal, Part 1, vol. 260, Sept. 1, 1982, p. 371-385. Research
  supported by the University of California}\ }\textbf {\bibinfo {volume}
  {260}},\ \bibinfo {pages} {371} (\bibinfo {year} {1982})}\BibitemShut
  {NoStop}%
\bibitem [{\citenamefont {Duncan}\ and\ \citenamefont
  {Thompson}(1992)}]{duncan1992magnetar}%
  \BibitemOpen
  \bibfield  {author} {\bibinfo {author} {\bibfnamefont {R.~C.}\ \bibnamefont
  {Duncan}}\ and\ \bibinfo {author} {\bibfnamefont {C.}~\bibnamefont
  {Thompson}},\ }\bibfield  {title} {\bibinfo {title} {Formation of very
  strongly magnetized neutron stars-implications for gamma-ray bursts},\
  }\href@noop {} {\bibfield  {journal} {\bibinfo  {journal} {Astrophysical
  Journal, Part 2-Letters (ISSN 0004-637X), vol. 392, no. 1, June 10, 1992, p.
  L9-L13. Research supported by NSERC.}\ }\textbf {\bibinfo {volume} {392}},\
  \bibinfo {pages} {L9} (\bibinfo {year} {1992})}\BibitemShut {NoStop}%
\bibitem [{\citenamefont {Gralla}\ \emph {et~al.}(2017)\citenamefont {Gralla},
  \citenamefont {Lupsasca},\ and\ \citenamefont
  {Philippov}}]{gralla2017quadrupole}%
  \BibitemOpen
  \bibfield  {author} {\bibinfo {author} {\bibfnamefont {S.~E.}\ \bibnamefont
  {Gralla}}, \bibinfo {author} {\bibfnamefont {A.}~\bibnamefont {Lupsasca}},\
  and\ \bibinfo {author} {\bibfnamefont {A.}~\bibnamefont {Philippov}},\
  }\bibfield  {title} {\bibinfo {title} {Inclined pulsar magnetospheres in
  general relativity: polar caps for the dipole, quadrudipole, and beyond},\
  }\href@noop {} {\bibfield  {journal} {\bibinfo  {journal} {The Astrophysical
  Journal}\ }\textbf {\bibinfo {volume} {851}},\ \bibinfo {pages} {137}
  (\bibinfo {year} {2017})}\BibitemShut {NoStop}%
\bibitem [{\citenamefont {Kalapotharakos}\ \emph {et~al.}(2021)\citenamefont
  {Kalapotharakos}, \citenamefont {Wadiasingh}, \citenamefont {Harding},\ and\
  \citenamefont {Kazanas}}]{kalapotharakos2021quadrupole}%
  \BibitemOpen
  \bibfield  {author} {\bibinfo {author} {\bibfnamefont {C.}~\bibnamefont
  {Kalapotharakos}}, \bibinfo {author} {\bibfnamefont {Z.}~\bibnamefont
  {Wadiasingh}}, \bibinfo {author} {\bibfnamefont {A.~K.}\ \bibnamefont
  {Harding}},\ and\ \bibinfo {author} {\bibfnamefont {D.}~\bibnamefont
  {Kazanas}},\ }\bibfield  {title} {\bibinfo {title} {The multipolar magnetic
  field of the millisecond pulsar psr j0030+ 0451},\ }\href@noop {} {\bibfield
  {journal} {\bibinfo  {journal} {The Astrophysical Journal}\ }\textbf
  {\bibinfo {volume} {907}},\ \bibinfo {pages} {63} (\bibinfo {year}
  {2021})}\BibitemShut {NoStop}%
\bibitem [{\citenamefont {Petri}(2015)}]{petri2015quadrupole}%
  \BibitemOpen
  \bibfield  {author} {\bibinfo {author} {\bibfnamefont {J.}~\bibnamefont
  {Petri}},\ }\bibfield  {title} {\bibinfo {title} {Multipolar electromagnetic
  fields around neutron stars: exact vacuum solutions and related properties},\
  }\href@noop {} {\bibfield  {journal} {\bibinfo  {journal} {Monthly Notices of
  the Royal Astronomical Society}\ }\textbf {\bibinfo {volume} {450}},\
  \bibinfo {pages} {714} (\bibinfo {year} {2015})}\BibitemShut {NoStop}%
\bibitem [{\citenamefont {P{\'e}tri}(2016)}]{petri2016quadrupole}%
  \BibitemOpen
  \bibfield  {author} {\bibinfo {author} {\bibfnamefont {J.}~\bibnamefont
  {P{\'e}tri}},\ }\bibfield  {title} {\bibinfo {title} {Radiation from an
  off-centred rotating dipole in vacuum},\ }\href@noop {} {\bibfield  {journal}
  {\bibinfo  {journal} {Monthly Notices of the Royal Astronomical Society}\
  }\textbf {\bibinfo {volume} {463}},\ \bibinfo {pages} {1240} (\bibinfo {year}
  {2016})}\BibitemShut {NoStop}%
\bibitem [{\citenamefont {Philippov}\ and\ \citenamefont
  {Kramer}(2022)}]{philippov2022pulsar_plasma}%
  \BibitemOpen
  \bibfield  {author} {\bibinfo {author} {\bibfnamefont {A.}~\bibnamefont
  {Philippov}}\ and\ \bibinfo {author} {\bibfnamefont {M.}~\bibnamefont
  {Kramer}},\ }\bibfield  {title} {\bibinfo {title} {Pulsar magnetospheres and
  their radiation},\ }\href@noop {} {\bibfield  {journal} {\bibinfo  {journal}
  {Annual Review of Astronomy and Astrophysics}\ }\textbf {\bibinfo {volume}
  {60}},\ \bibinfo {pages} {495} (\bibinfo {year} {2022})}\BibitemShut
  {NoStop}%
\bibitem [{\citenamefont {Zhang}(2022)}]{zhang2022Compton}%
  \BibitemOpen
  \bibfield  {author} {\bibinfo {author} {\bibfnamefont {B.}~\bibnamefont
  {Zhang}},\ }\bibfield  {title} {\bibinfo {title} {Coherent inverse compton
  scattering by bunches in fast radio bursts},\ }\href@noop {} {\bibfield
  {journal} {\bibinfo  {journal} {The Astrophysical Journal}\ }\textbf
  {\bibinfo {volume} {925}},\ \bibinfo {pages} {53} (\bibinfo {year}
  {2022})}\BibitemShut {NoStop}%
\bibitem [{\citenamefont {Kaspi}\ and\ \citenamefont
  {Beloborodov}(2017)}]{kaspi2017magnetars}%
  \BibitemOpen
  \bibfield  {author} {\bibinfo {author} {\bibfnamefont {V.~M.}\ \bibnamefont
  {Kaspi}}\ and\ \bibinfo {author} {\bibfnamefont {A.~M.}\ \bibnamefont
  {Beloborodov}},\ }\bibfield  {title} {\bibinfo {title} {Magnetars},\
  }\href@noop {} {\bibfield  {journal} {\bibinfo  {journal} {Annual Review of
  Astronomy and Astrophysics}\ }\textbf {\bibinfo {volume} {55}},\ \bibinfo
  {pages} {261} (\bibinfo {year} {2017})}\BibitemShut {NoStop}%
\bibitem [{\citenamefont {Lu}\ \emph {et~al.}(2020)\citenamefont {Lu},
  \citenamefont {Kumar},\ and\ \citenamefont {Zhang}}]{lu2020FRB}%
  \BibitemOpen
  \bibfield  {author} {\bibinfo {author} {\bibfnamefont {W.}~\bibnamefont
  {Lu}}, \bibinfo {author} {\bibfnamefont {P.}~\bibnamefont {Kumar}},\ and\
  \bibinfo {author} {\bibfnamefont {B.}~\bibnamefont {Zhang}},\ }\bibfield
  {title} {\bibinfo {title} {A unified picture of galactic and cosmological
  fast radio bursts},\ }\href@noop {} {\bibfield  {journal} {\bibinfo
  {journal} {Monthly Notices of the Royal Astronomical Society}\ }\textbf
  {\bibinfo {volume} {498}},\ \bibinfo {pages} {1397} (\bibinfo {year}
  {2020})}\BibitemShut {NoStop}%
\bibitem [{\citenamefont {Wadiasingh}\ and\ \citenamefont
  {Timokhin}(2019)}]{wadiasingh2019FRB}%
  \BibitemOpen
  \bibfield  {author} {\bibinfo {author} {\bibfnamefont {Z.}~\bibnamefont
  {Wadiasingh}}\ and\ \bibinfo {author} {\bibfnamefont {A.}~\bibnamefont
  {Timokhin}},\ }\bibfield  {title} {\bibinfo {title} {Repeating fast radio
  bursts from magnetars with low magnetospheric twist},\ }\href@noop {}
  {\bibfield  {journal} {\bibinfo  {journal} {The Astrophysical Journal}\
  }\textbf {\bibinfo {volume} {879}},\ \bibinfo {pages} {4} (\bibinfo {year}
  {2019})}\BibitemShut {NoStop}%
\bibitem [{\citenamefont {Archer}\ \emph {et~al.}(2018)\citenamefont {Archer},
  \citenamefont {Benbow}, \citenamefont {Bird}, \citenamefont {Brose},
  \citenamefont {Buchovecky}, \citenamefont {Buckley}, \citenamefont {Bugaev},
  \citenamefont {Connolly}, \citenamefont {Cui}, \citenamefont {Daniel} \emph
  {et~al.}}]{archer2018TeV_electrons}%
  \BibitemOpen
  \bibfield  {author} {\bibinfo {author} {\bibfnamefont {A.}~\bibnamefont
  {Archer}}, \bibinfo {author} {\bibfnamefont {W.}~\bibnamefont {Benbow}},
  \bibinfo {author} {\bibfnamefont {R.}~\bibnamefont {Bird}}, \bibinfo {author}
  {\bibfnamefont {R.}~\bibnamefont {Brose}}, \bibinfo {author} {\bibfnamefont
  {M.}~\bibnamefont {Buchovecky}}, \bibinfo {author} {\bibfnamefont
  {J.}~\bibnamefont {Buckley}}, \bibinfo {author} {\bibfnamefont
  {V.}~\bibnamefont {Bugaev}}, \bibinfo {author} {\bibfnamefont
  {M.}~\bibnamefont {Connolly}}, \bibinfo {author} {\bibfnamefont
  {W.}~\bibnamefont {Cui}}, \bibinfo {author} {\bibfnamefont {M.}~\bibnamefont
  {Daniel}}, \emph {et~al.},\ }\bibfield  {title} {\bibinfo {title}
  {Measurement of cosmic-ray electrons at tev energies by veritas},\
  }\href@noop {} {\bibfield  {journal} {\bibinfo  {journal} {Physical Review
  D}\ }\textbf {\bibinfo {volume} {98}},\ \bibinfo {pages} {062004} (\bibinfo
  {year} {2018})}\BibitemShut {NoStop}%
\bibitem [{\citenamefont {Recchia}\ \emph {et~al.}(2019)\citenamefont
  {Recchia}, \citenamefont {Gabici}, \citenamefont {Aharonian},\ and\
  \citenamefont {Vink}}]{recchia2019TeV_electrons}%
  \BibitemOpen
  \bibfield  {author} {\bibinfo {author} {\bibfnamefont {S.}~\bibnamefont
  {Recchia}}, \bibinfo {author} {\bibfnamefont {S.}~\bibnamefont {Gabici}},
  \bibinfo {author} {\bibfnamefont {F.}~\bibnamefont {Aharonian}},\ and\
  \bibinfo {author} {\bibfnamefont {J.}~\bibnamefont {Vink}},\ }\bibfield
  {title} {\bibinfo {title} {Local fading accelerator and the origin of tev
  cosmic ray electrons},\ }\href@noop {} {\bibfield  {journal} {\bibinfo
  {journal} {Physical Review D}\ }\textbf {\bibinfo {volume} {99}},\ \bibinfo
  {pages} {103022} (\bibinfo {year} {2019})}\BibitemShut {NoStop}%
\bibitem [{\citenamefont {Aab}\ \emph {et~al.}(2020)\citenamefont {Aab},
  \citenamefont {Abreu}, \citenamefont {Aglietta}, \citenamefont {Albury},
  \citenamefont {Allekotte}, \citenamefont {Almela}, \citenamefont {Castillo},
  \citenamefont {Alvarez-Mu{\~n}iz}, \citenamefont {Batista}, \citenamefont
  {Anastasi} \emph {et~al.}}]{aab2020rays}%
  \BibitemOpen
  \bibfield  {author} {\bibinfo {author} {\bibfnamefont {A.}~\bibnamefont
  {Aab}}, \bibinfo {author} {\bibfnamefont {P.}~\bibnamefont {Abreu}}, \bibinfo
  {author} {\bibfnamefont {M.}~\bibnamefont {Aglietta}}, \bibinfo {author}
  {\bibfnamefont {J.~M.}\ \bibnamefont {Albury}}, \bibinfo {author}
  {\bibfnamefont {I.}~\bibnamefont {Allekotte}}, \bibinfo {author}
  {\bibfnamefont {A.}~\bibnamefont {Almela}}, \bibinfo {author} {\bibfnamefont
  {J.~A.}\ \bibnamefont {Castillo}}, \bibinfo {author} {\bibfnamefont
  {J.}~\bibnamefont {Alvarez-Mu{\~n}iz}}, \bibinfo {author} {\bibfnamefont
  {R.~A.}\ \bibnamefont {Batista}}, \bibinfo {author} {\bibfnamefont {G.~A.}\
  \bibnamefont {Anastasi}}, \emph {et~al.},\ }\bibfield  {title} {\bibinfo
  {title} {Features of the energy spectrum of cosmic rays above 2.5$\times$ 10
  18 ev using the pierre auger observatory},\ }\href@noop {} {\bibfield
  {journal} {\bibinfo  {journal} {Physical review letters}\ }\textbf {\bibinfo
  {volume} {125}},\ \bibinfo {pages} {121106} (\bibinfo {year}
  {2020})}\BibitemShut {NoStop}%
\bibitem [{\citenamefont {Drury}(1994)}]{drury1994Rays}%
  \BibitemOpen
  \bibfield  {author} {\bibinfo {author} {\bibfnamefont {L.~O.}\ \bibnamefont
  {Drury}},\ }\bibfield  {title} {\bibinfo {title} {Acceleration of cosmic
  rays},\ }\href@noop {} {\bibfield  {journal} {\bibinfo  {journal}
  {Contemporary Physics}\ }\textbf {\bibinfo {volume} {35}},\ \bibinfo {pages}
  {231} (\bibinfo {year} {1994})}\BibitemShut {NoStop}%
\bibitem [{\citenamefont {Blackburn}\ \emph {et~al.}(2017)\citenamefont
  {Blackburn}, \citenamefont {Ilderton}, \citenamefont {Murphy},\ and\
  \citenamefont {Marklund}}]{blackburn2017scaling}%
  \BibitemOpen
  \bibfield  {author} {\bibinfo {author} {\bibfnamefont {T.}~\bibnamefont
  {Blackburn}}, \bibinfo {author} {\bibfnamefont {A.}~\bibnamefont {Ilderton}},
  \bibinfo {author} {\bibfnamefont {C.}~\bibnamefont {Murphy}},\ and\ \bibinfo
  {author} {\bibfnamefont {M.}~\bibnamefont {Marklund}},\ }\bibfield  {title}
  {\bibinfo {title} {Scaling laws for positron production in
  laser--electron-beam collisions},\ }\href@noop {} {\bibfield  {journal}
  {\bibinfo  {journal} {Physical Review A}\ }\textbf {\bibinfo {volume} {96}},\
  \bibinfo {pages} {022128} (\bibinfo {year} {2017})}\BibitemShut {NoStop}%
\end{thebibliography}%

\end{document}